%% file: main_techreport.tex
\documentclass[]{atomflow}

\usepackage{booktabs}            % professional-quality tables
\usepackage{multirow}            % tabular cells spanning multiple rows
\usepackage{amsmath, amsfonts, amsthm, amssymb, bbm} % blackboard math symbols
\usepackage{graphicx}            % figures
\usepackage{duckuments}          % sample images
\usepackage{soul}                % highlighting text
\usepackage{titlesec}
\usepackage[toc,page]{appendix}  % apHybrpendix
\usepackage{titletoc}            % table of contents in appendix
\usepackage{tikz}
\usepackage{lineno}              % line numbers
\usepackage{marvosym}
\usepackage{fontawesome}
\usepackage{listings}
\usepackage{hyperref}

\usepackage[version=4]{mhchem}   % chemical symbols

\usepackage{mathptmx}
\usepackage{mathrsfs}
\DeclareMathAlphabet{\mathcal}{OMS}{cmsy}{m}{n}  % distinguish 区分\mathcal,\mathscr
\usepackage{ulem} % strikethrough 删除线 \sout
\usepackage{xpatch}
\usepackage{url}
\usepackage{texpower}
\usepackage{makecell}
\usepackage{colortbl}
\usepackage{subcaption}

\definecolor{tableHead}{RGB}{216,214,194}
\definecolor{tableContent}{RGB}{235,234,222}
\usepackage{tabularx}

\usepackage{tcolorbox}
\definecolor{backred}{RGB}{255, 190, 190}
\definecolor{backblue}{RGB}{220, 230, 250}

\usepackage{marvosym}
\usepackage{lineno}

\definecolor{modification}{RGB}{0,0,0}  % black

\definecolor{cred}{HTML}{FF6B6B}
\definecolor{cyellow}{HTML}{FEC260}
\definecolor{cgreen}{HTML}{6BCB77}
\definecolor{cgreen}{HTML}{70AD47}
\definecolor{cblue}{HTML}{4D96FF}
\definecolor{cpurple}{HTML}{2A0944}
\definecolor{ggray}{RGB}{127,127,127}
\definecolor{aliceblue}{rgb}{0.94, 0.97, 1.0}

\definecolor{aliceblue}{rgb}{0.94,0.97,1.0}
\definecolor{darkgreen}{RGB}{25,25,180}

\definecolor{myblue}{HTML}{4B86E8}
\definecolor{myorange2}{HTML}{FE9A00}
\definecolor{mygreen}{HTML}{097969}

\usepackage{makecell}   % 表格内换行

\newcommand{\codeinline}[1]{%
  \colorbox{gray!10}{\lstinline[basicstyle=\ttfamily\small]|#1|}}

\input{seed/macro}
\title{MoleCode unlocks structural intelligence\\in large language models}

\author[1,2,*]{Zhiyuan Yan}
\author[2,*]{Chen Liu}
\author[1,2,*]{Boxuan Zhao}
\author[2]{Kaiqing Lin}
\author[2]{Jixiang Zhao}
\author[1,2]{Yimi Wang}
\author[1]{Liuzhenghao Lv}
\author[1]{Hao Li}
\author[2]{Shanzhuo Zhang}
\author[1\textsuperscript{$\dag$}]{Li Yuan}
\author[1\textsuperscript{$\dag$}]{Fanyang Mo}

\affiliation[1]{Peking University Shenzhen Graduage School}
\affiliation[2]{AtomFlow}
\affiliation[*]{Equal Contribution}
\affiliation[\textsuperscript{$\dag$}]{Correspondence: \url{yuanli-ece@pku.edu.cn}, \url{fmo@pku.edu.cn}}

\checkdata[Project Page]{\href{https://atomflow-ai.com}{\url{https://atomflow-ai.com}}}

\begin{document}

\abstract{
\input{abstract}
}

\maketitle

\input{all_content}
\end{document}

%% file: seed/macro.tex
\usepackage{natbib}
\usepackage{latexsym}

\usepackage{url}
\usepackage{amssymb}
\usepackage[utf8]{inputenc}
\usepackage{microtype}
\usepackage{booktabs}
\usepackage{pifont} 
\usepackage{multirow}
\usepackage{makecell}
\usepackage{paralist}
\usepackage{xspace}
\usepackage{color}
\usepackage{xcolor}
\usepackage{colortbl}
\usepackage{hyperref} 
\usepackage[edges]{forest}
\usepackage{tikz} 
\usepackage{caption}
\usepackage{amsfonts}

\hypersetup{
    colorlinks,
    linkcolor={blue!80!black},
    citecolor={blue!80!black},
}
\tikzset{
    root/.style =             {align=center, text width=1cm, rounded corners=3pt, line width=0.3mm, fill=gray!10, draw=gray!80, font=\small},
    demographic/.style =         {align=center, text width=1.8cm, rounded corners=3pt, line width=0.3mm, fill=blue!10, draw=blue!80, font=\footnotesize},
    demographic_work/.style =    {align=center, text width=10cm, rounded corners=3pt, line width=0.3mm, fill=blue!10, draw=blue!0, font=\footnotesize},
    character/.style =         {align=center, text width=1.8cm, rounded corners=3pt, line width=0.3mm, fill=red!10, draw=red!80, font=\footnotesize},
    character_work/.style =    {align=center, text width=10cm, rounded corners=3pt, line width=0.3mm, fill=red!10, draw=red!0, font=\footnotesize},
    personalization/.style =           {align=center, text width=1.8cm, rounded corners=3pt, line width=0.3mm, fill=cyan!10, draw=cyan!80, font=\footnotesize},
    personalization_work/.style =      {align=center, text width=10cm, rounded corners=3pt, line width=0.3mm, fill=cyan!10, draw=cyan!0, font=\footnotesize},
    risk/.style =         {align=center, text width=1.8cm, rounded corners=3pt, line width=0.3mm, fill=orange!10, draw=orange!80, font=\footnotesize},
    risk_work/.style =    {align=center, text width=10cm, rounded corners=3pt, line width=0.3mm, fill=orange!10, draw=orange!0, font=\footnotesize},
}

\usepackage{CJK}

%% file: abstract.tex
Molecules are graphs, but large language models~(LLMs) are usually asked to reason about them through linear strings. The most popular molecular representation, SMILES, compresses atoms, bonds, branches and rings into a compact sequence in which topology is implicit, forcing LLMs to reconstruct molecular structure before performing the requested chemical operation. Here we introduce \textbf{MoleCode}, an LLM-native, training-free, graph-explicit molecular language in which all molecular components are represented as typed entities with persistent identifiers and explicit relations. \textbf{MoleCode} makes molecular topology directly readable, editable and auditable within the language context, allowing an LLM to operate on structure rather than recover it from syntax. Across molecular reasoning, editing, generation and analysis tasks, this representational shift improves frontier LLMs most strongly when structural access is limiting: unfamiliar molecules, topology-sensitive operations, larger structures and repetitive polymers. It also changes how inference is allocated, replacing long reasoning traces devoted to implicit structural reconstruction with shorter, more chemically directed reasoning over explicit atoms and bonds. In molecular optimization, this enables localized, property-aligned edits that preserve structural similarity to the starting compounds. The same Subgraph--Node--Edge grammar extends beyond small molecules to polymers, Markush structures, mechanism-style transformations and interleaved scientific documents, including research articles and patent disclosures in which chemical information is distributed across text and images. These results suggest that the interface between scientific objects and LLMs should not treat structure as something to be decoded from text. When the object of reasoning is relational, the structure itself should be part of the language.

%% file: all_content.tex
\begin{figure*}[!ht]
\centering
\includegraphics[width=\textwidth]{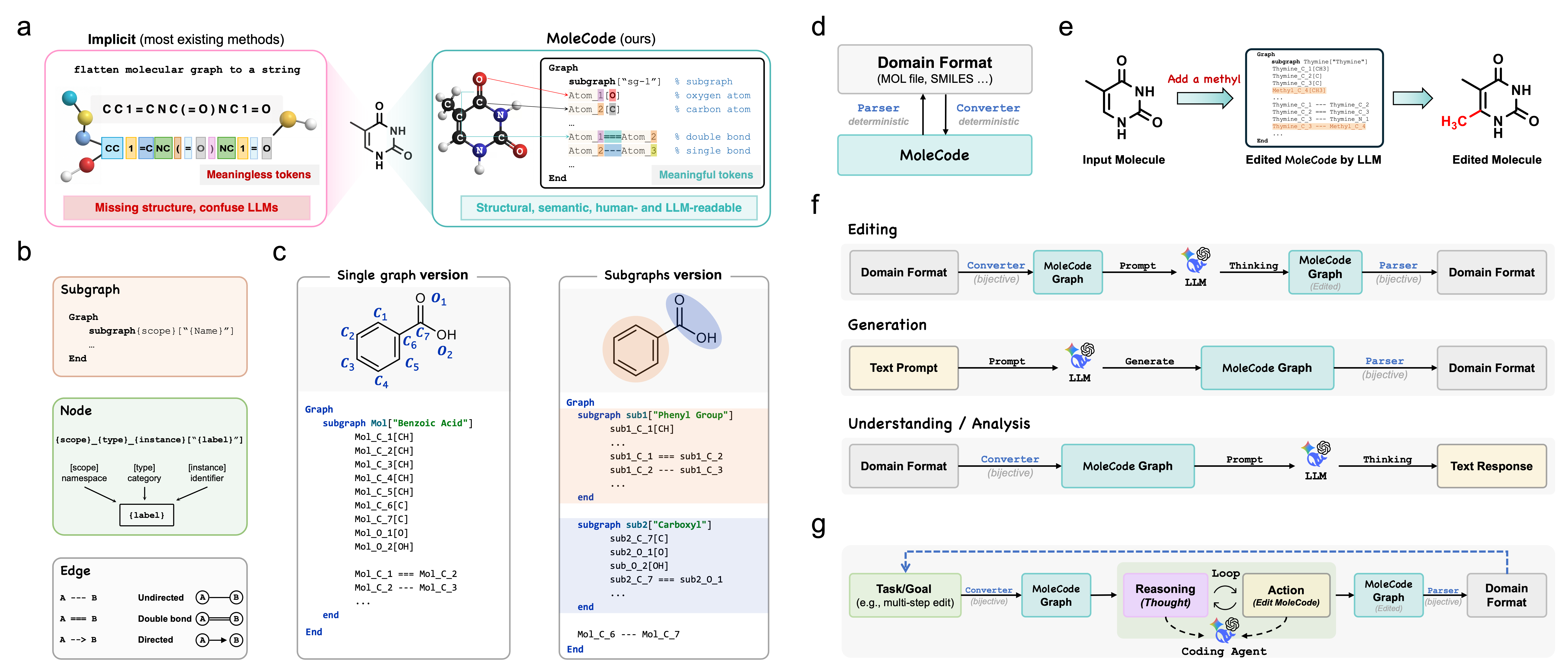}
\caption{
\small
\textbf{MoleCode makes structure explicit, editable, and operable for LLMs.}
\textbf{a}, Implicit and explicit representations of the same molecule. Conventional molecular strings such as SMILES flatten molecular graphs into semantically arbitrary token sequences in which topology is implicit and must be reconstructed during reasoning. MoleCode instead encodes atoms and bonds as explicit graph representations with persistent identifiers and readable structural semantics.
\textbf{b}, MoleCode abstract grammar. MoleCode is built mainly from three primitives: \texttt{Subgraph}, \texttt{Node}, and \texttt{Edge}. Subgraphs define structural scopes, nodes encode typed entities with persistent identifiers, and edges encode explicit relations between nodes.
\textbf{c}, Implementation examples of MoleCode. A complex molecule can be represented either as a single graph or as multiple subgraphs while preserving exact topology and deterministic equivalence.
\textbf{d}, Deterministic bidirectional conversion between domain formats and MoleCode. Existing molecular representations, including MOL files and SMILES, can be converted to and from MoleCode in a deterministic manner without loss of structural information.
\textbf{e}, Structure-aware molecular manipulation through explicit graph operations. A frontier LLM edits a molecule directly in MoleCode space by performing localized graph modifications.
\textbf{f}, Standardized LLM workflows using MoleCode. MoleCode supports molecular editing, generation, understanding, and analysis within a unified graph-language interface, enabling frontier LLMs to operate over explicit topology.
\textbf{g}, Agentic long-task workflow enabled by MoleCode. For multi-step molecular tasks, an LLM-based coding agent can iteratively reason over a MoleCode graph and perform explicit graph-level actions. %See Supplementary Fig.~\ref{fig:supp_agentic} for details.
}
\label{fig:concept}
\end{figure*}

\section*{Main}

Large language models~(LLMs) are emerging as general-purpose interfaces for molecular science~\cite{ashyrmamatov2025survey, bhattacharya2024large, park2024llamo}. Beyond predicting molecular properties, they can interpret natural-language instructions, answer chemical questions, propose molecular edits, reason over design constraints and coordinate external tools in multi-step workflows~\cite{ChemBench, molt5, pei2023biot5, lv2024navigating, liu2023molca, molinstructions, cao2025instructmol, seidl2023clamp1, li20243dmolm, yu2024llasmol}. This shift changes the role of molecular representation: a molecule is no longer only an input to a specialized predictor~\cite{givechian2026immunostruct}, but rather becomes an object that the model must read, inspect, modify and communicate about through language~\cite{molinstructions, cao2025instructmol}.

This creates a fundamental interface problem. Molecules are graphs: atoms are nodes, bonds are edges, and chemical behaviour depends on topology~\cite{reiser2022graph, skinnider2024invalid}. However, the dominant text-based molecular representation, \textit{Simplified Molecular Input Line Entry System}~(SMILES)~\cite{SMILES}, presents this graph as a one-dimensional string~\cite{jang2025improving}. In SMILES, connectivity is implicit, branches are encoded through syntax and ring closures are represented by positional indices~\cite{guo2023gpt_chemistry, ChemBench, hao2025detect, ChemCoTBench}. Thus, when an LLM is asked to predict the molecular formula, recognize a functional group, make a local edit or satisfy a structural constraint, it must first internally reconstruct the molecular graph before it can reason about the requested operation~\cite{liu2026fgbench, wu2025molerr2fix, liu2023molxpt, zhong2022root}.

This bottleneck reflects a broader principle: representations shape what kinds of reasoning are easy~\cite{zhang1994representations, zhang1997nature}. Arabic numerals with positional notation made arithmetic scalable by turning magnitude and place into explicit symbols~\cite{smith1911hindu}. The structural formulas transformed chemistry by making explicit representations of the molecular composition and connectivity~\cite{olmsted1997chemistry, talanquer2022complexity}. For LLM-based molecular science, the analogous question is whether molecular topology is directly available as part of the language interface, or hidden behind syntax that the model must decode~\cite{jang2025improving, ChemBench}.

Existing molecular representations and models only partially address this issue. SELFIES~\cite{SELFIES} and T-SMILES~\cite{t_SMILES} improve syntactic validity and robustness, but still present molecules as sequential strings rather than explicit node-edge declarations. Graph neural networks~\cite{gilmer2017neural, kipf2017semi, schutt2018schnet, reiser2022graph} operate directly on molecular topology and have achieved strong performance in property prediction and molecular representation learning~\cite{zhouunimol, hafidi2020graphcl, kv-plm}. However, they are typically specialized predictors rather than general interfaces for open-ended molecular reasoning, natural-language constraints and iterative editing~\cite{reiser2022graph, cai2025mollangbench, graphmvp, lyy-blending2d3d, lyy-unicorn2d3d}. Hybrid graph-language systems attempt to connect these regimes by projecting molecular graphs into vectors or textual sequences~\cite{molt5, liu2023molca, momu, kim2026mol-llama}, but this compression weakens locality, editability and traceability. Thus, current approaches separate two capabilities that should ideally coexist: accurate structural perception and flexible language-based reasoning.

To address these challenges, we introduce MoleCode, an LLM-native, training-free, graph-explicit molecular language designed to make molecular topology directly operable by LLMs without modifying model weights. In MoleCode, atoms and bonds are written as typed declarations with persistent identifiers, so connectivity is stated directly rather than inferred from sequential syntax~(Fig.~\ref{fig:concept}a). MoleCode is built from a simple Subgraph--Node--Edge grammar that supports both single graph serializations and hierarchical graph decompositions~(Fig.~\ref{fig:concept}b-c). Standard molecular formats, including MOL files and SMILES, can be converted into MoleCode and reconstructed through deterministic bidirectional pipelines~(Fig.~\ref{fig:concept}d). Because atoms and bonds are explicit textual objects, chemically meaningful edits become localized graph operations, such as adding a methyl group through the addition of one node and one edge~(Fig.~\ref{fig:concept}e). The same interface supports molecular editing, generation, understanding and analysis~(Fig.~\ref{fig:concept}f), and naturally integrates into multi-step agentic workflows in which a coding agent iteratively reasons over, edits and validates graph objects~(Fig.~\ref{fig:concept}g).

This representation changes the role of the LLMs. Under implicit molecular strings, graph structures must first be inferred before chemical reasoning can begin. Under MoleCode, topology is already present in the context window as readable structure, allowing the model to operate directly on atoms, bonds, subgraphs and transformations. We therefore hypothesized that MoleCode would be most beneficial in settings where structural access is limiting, including unfamiliar molecules, topology-sensitive tasks and goal-directed edits requiring chemically localized modifications.

We tested this hypothesis across small-molecule reasoning, molecular optimization, inference-cost analysis, polymers, and higher-order chemical structure. MoleCode improved molecular reasoning most strongly when structural generalization rather than memorization was required~(Fig.~\ref{fig:results_1_main}). It enabled more coherent molecular optimization through localized, property-aligned edits~(Fig.~\ref{fig:results_2_chemistry}), and reallocated inference from structural reconstruction towards more chemically directed reasoning~(Fig.~\ref{fig:results_3_scaling}). The benefits became stronger for long and repetitive polymers, where implicit sequential encodings degraded rapidly~(Fig.~\ref{fig:results_4_long_molecules}). Finally, the same graph-native abstraction extended naturally to Markush structures, mechanism-style transformations and interleaved image-text chemistry documents, including research articles and patent disclosures~(Fig.~\ref{fig:results_5_extension}). Together, these results support a broader principle for scientific language interfaces: when the object of reasoning is structured, the structure itself should be directly accessible to the LLM, rather than hidden behind implicit sequential syntax.

\section*{Results}

\subsection*{MoleCode improves molecular reasoning through structural generalization}

\begin{figure*}[!th]
\centering
\includegraphics[width=\textwidth]{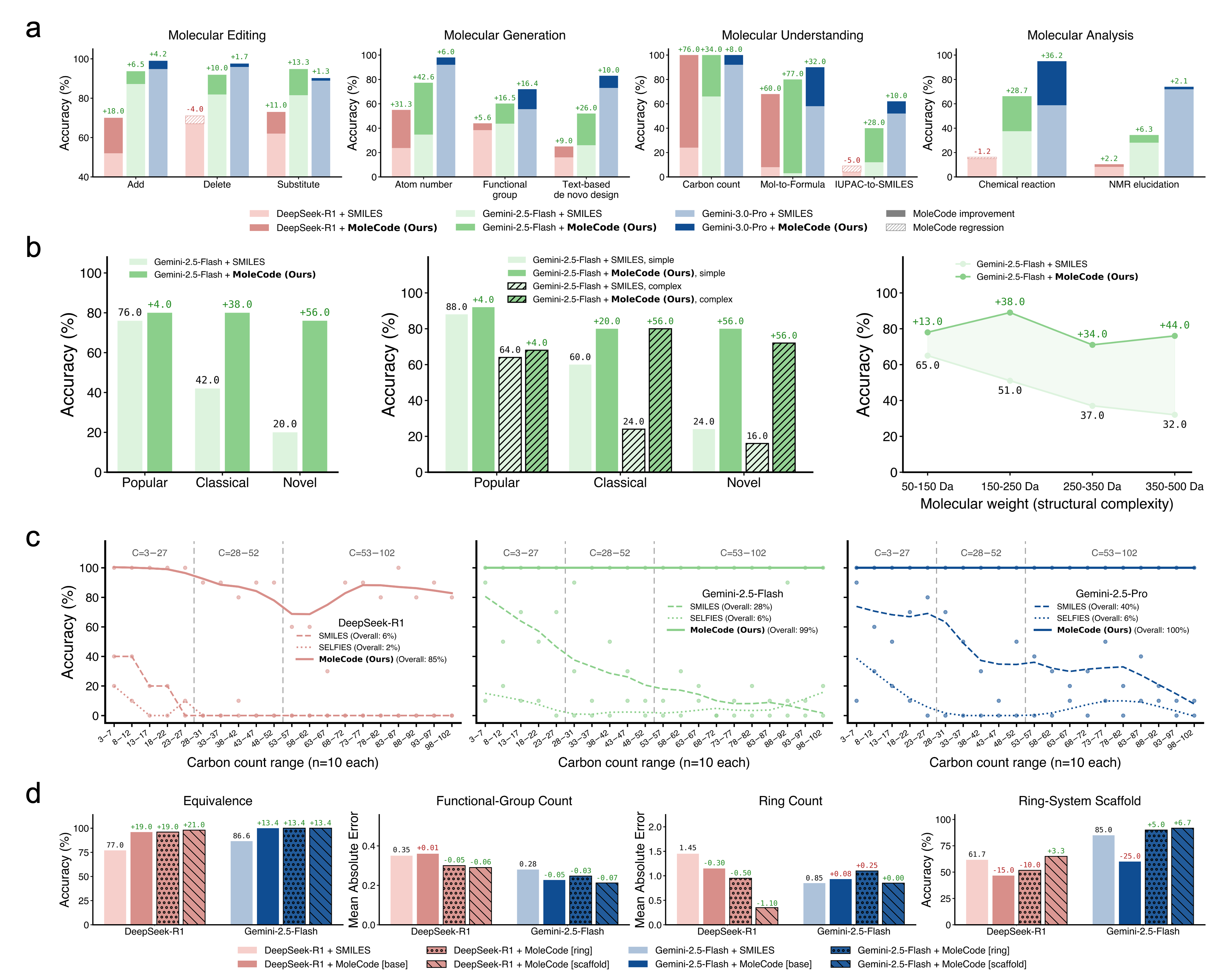}
\caption{
\small
\textbf{Explicit structure improves molecular reasoning by supporting structural generalization rather than memorization.}
\textbf{a}, Performance comparison across four molecular task families: editing, generation, understanding, and analysis. Across three frontier LLMs, MoleCode consistently improves over SMILES, with the largest gains on structure-intensive tasks such as atom-number-constrained generation, molecular formula prediction, carbon counting, and chemical reaction prediction.
\textbf{b}, Generalization across molecular familiarity and structural complexity. Left, SMILES performs well only on popular (i.e., frequently seen) molecules but degrades when encountering classical and especially novel molecules, whereas MoleCode remains highly consistent. Middle, stratification by familiarity and complexity shows that SMILES collapses most strongly on structurally complex and unfamiliar molecules, while MoleCode remains robust. Right, as molecular weight increases, SMILES accuracy declines monotonically, whereas MoleCode preserves high accuracy.
\textbf{c}, Generalization across molecular size. SMILES and SELFIES degrade as molecular size increases, while MoleCode remains substantially more stable, indicating that explicit topology improves reasoning under increasing structural complexity.
\textbf{d}, Representation ablation on molecular understanding tasks. MoleCode variants with increasingly explicit structural annotations generally improve over SMILES, with scaffold-aware encoding providing the strongest and most consistent performance across models and tasks.
}
\label{fig:results_1_main}
\end{figure*}

We first asked whether making molecular topology explicit improves LLM reasoning across different forms of molecular work. We evaluated three frontier LLMs, DeepSeek-R1~\cite{guo2025deepseek}, Gemini-2.5-Flash~\cite{gemini2.5} and Gemini-3-Pro~\cite{gemini3}, on a unified benchmark spanning four task families: molecular editing, molecular generation, molecular understanding and molecular analysis~(Fig.~\ref{fig:results_1_main}a). These tasks ranged from local graph operations, including atom addition, deletion and substitution, to structure-sensitive problems such as atom-number-constrained generation, molecular formula prediction, carbon counting, IUPAC-to-SMILES conversion, reaction prediction and nuclear magnetic resonance elucidation.

Across all four task families, replacing SMILES with MoleCode consistently improved performance~(Fig.~\ref{fig:results_1_main}a). The largest gains appeared in tasks that require accurate access to molecular topology rather than recognition of local string patterns. For Gemini-3-Pro, reaction prediction accuracy increased from 58.8\% to 95.0\%, and molecular formula prediction increased from 58.0\% to 90.0\%. Gemini-2.5-Flash showed similarly large shifts on structure-sensitive generation and understanding tasks, with atom-number-constrained generation increasing from 34.7\% to 77.3\% and molecular formula prediction increasing from 3.0\% to 80.0\%. By contrast, the gap was smaller on local editing tasks, where only limited structural context must be inferred. This pattern suggests that the benefit of MoleCode grows when the model would otherwise need to reconstruct hidden topology from sequential syntax.

We next examined whether these gains reflected structural generalization rather than improved matching to familiar molecular strings. We stratified molecules into popular, classical and novel familiarity tiers using PubChem~\cite{kim2023pubchem} occurrence frequency as an approximate proxy for molecular familiarity and likely pretraining exposure~(Fig.~\ref{fig:results_1_main}b, left). Under SMILES, accuracy decreased sharply as molecules became less familiar, dropping from 42\% on classical molecules to 20\% on novel molecules. MoleCode remained substantially more stable, maintaining approximately 76--80\% accuracy across all three tiers. This stability is consistent with the interpretation that explicit topology helps models reason over molecular structure rather than rely primarily on familiar string forms.

We then asked whether this familiarity effect interacted with structural complexity. Joint stratification by molecular familiarity and complexity showed that SMILES degraded most strongly when molecules were both unfamiliar and structurally complex~(Fig.~\ref{fig:results_1_main}b, middle). On simple classical compounds, SMILES and MoleCode performed comparably. On complex novel molecules, however, SMILES accuracy decreased to 16\%, whereas MoleCode remained at 72\%. This pattern suggests that the limiting factor is not only chemical knowledge or prior exposure to familiar compounds, but also reliable access to topology when the structure is unfamiliar and difficult to recover from sequential syntax.

This representational bottleneck became more pronounced as molecular size increased. As molecular weight increased from 50--150 Da to 350--500 Da, SMILES accuracy decreased monotonically from 65\% to 32\%, whereas MoleCode remained comparatively stable from 78\% to 76\%~(Fig.~\ref{fig:results_1_main}b, right). The same trend appeared when scaling directly with carbon count across three molecular representations and three frontier LLMs~(Fig.~\ref{fig:results_1_main}c). SMILES and SELFIES progressively degraded with increasing molecular size, whereas MoleCode remained substantially more stable across the same complexity range. Thus, the value of explicit topology increases as the structural burden of the molecule grows.

To identify which aspects of MoleCode contributed to these improvements, we performed a representation ablation spanning progressively richer variants, including base, ring-aware and scaffold-aware encodings~(Fig.~\ref{fig:results_1_main}d). We evaluated these variants on molecular equivalence, functional-group counting, ring counting and ring-system scaffold identification. MoleCode variants improved over SMILES baselines across most tasks and models. The gains generally increased as structural annotations became richer, with scaffold-aware encoding producing the strongest and most consistent performance, particularly on scaffold-sensitive tasks such as ring-system identification. These ablations support the conclusion that the key factor is not prompt formatting alone, but rather the degree to which molecular topology is made explicit and accessible.

Together, these experiments reveal a consistent pattern across task families, familiarity regimes and complexity scales: MoleCode improves molecular reasoning most strongly when success depends on structural generalization rather than memorization of canonical molecular strings.

\subsection*{MoleCode enables chemically grounded molecular optimization}

\begin{figure*}[!th]
\centering
\includegraphics[width=\textwidth]{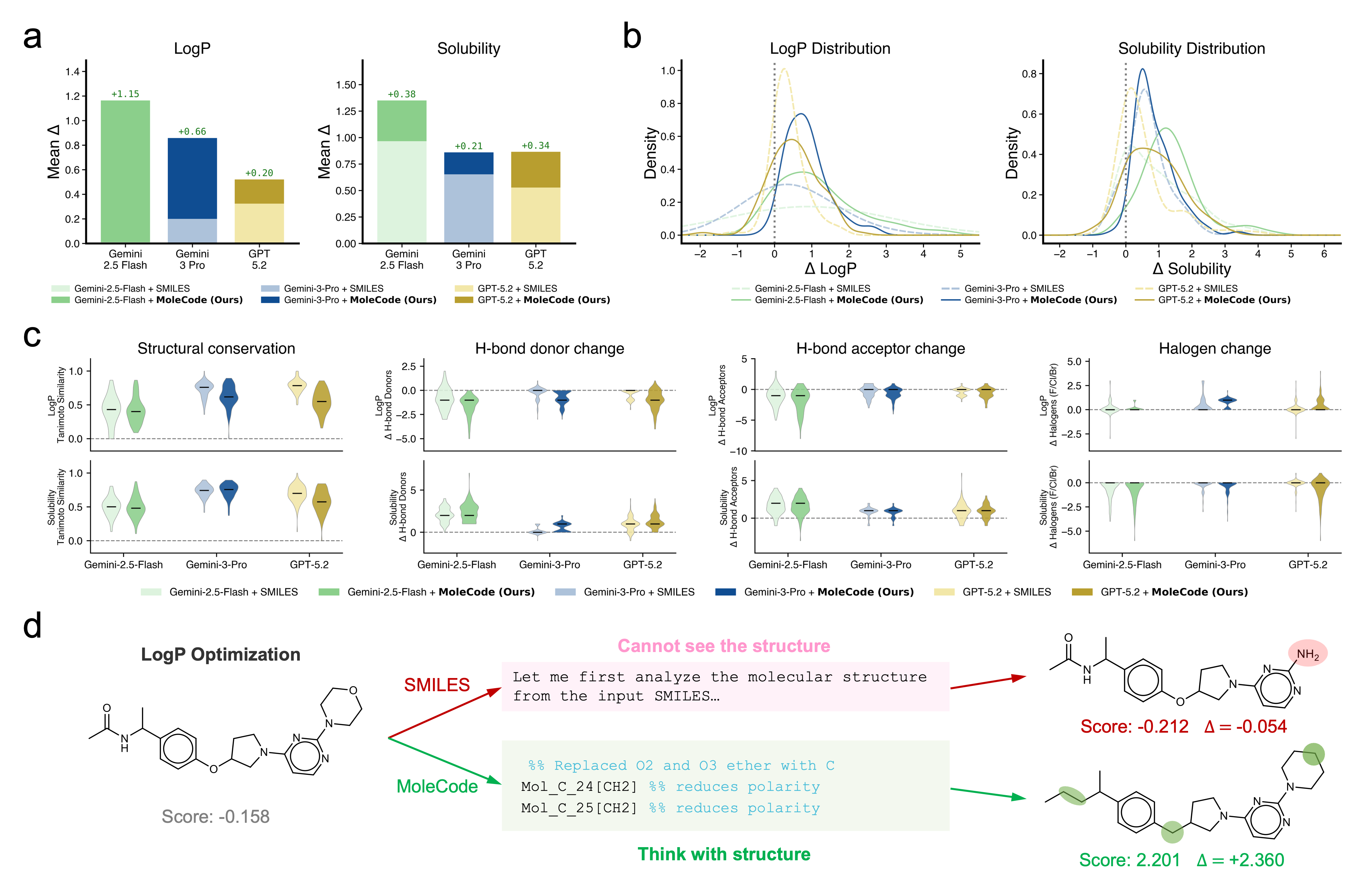}
\caption{
\small
\textbf{MoleCode improves goal-directed molecular design and induces chemically interpretable edits.}
\textbf{a}, Cross-model optimization gains for water-octanol
partition coefficient~(LogP) and solubility. Bars show mean property improvement~($\Delta$) from the starting molecules. MoleCode consistently improves optimization outcomes across both objectives and all three models.
\textbf{b}, Distribution of optimization gains. Kernel-density curves show that MoleCode shifts the distribution of $\Delta$LogP and $\Delta$Solubility toward larger improvements, indicating that the gains are broadly distributed rather than driven by a small number of outliers.
\textbf{c}, Descriptor-level analysis of the edits produced during optimization. For LogP optimization, MoleCode produces more task-aligned structural changes, including reduced hydrogen-bond donor and acceptor counts and increased halogenation. For solubility optimization, MoleCode shows increased hydrogen-bond donor tendency and reduced halogenation. Structural conservation is also tracked by Tanimoto similarity.
\textbf{d}, Representative LogP optimization example. Starting from the same molecule, SMILES fails to ground the edit in the structural context and produces a degraded candidate. MoleCode identifies specific graph nodes and performs chemically interpretable edits, including amide nitrogen modification and ether-to-methylene substitutions, yielding a substantially improved molecule. All references to ``LogP'' in this study denote ``penalized LogP'' as defined in prior works~\cite{kusner2017grammar, jin2018junction}.
}
\label{fig:results_2_chemistry}
\end{figure*}

Reasoning about molecular structure is only one requirement for molecular design. Practical molecular engineering also requires models to modify structure while preserving chemical coherence and aligning edits with a target objective. We therefore asked whether explicit graph structure also improves goal-directed molecular optimization.

We compared MoleCode and SMILES across Gemini-2.5-Flash, Gemini-3-Pro and GPT-5.2~\cite{GPT5} on ChemCoTBench~\cite{ChemCoTBench}, using two optimization objectives: lipophilicity measured by water-octanol partition coefficient~(LogP)~\cite{kusner2017grammar, jin2018junction}, and aqueous solubility. Across both objectives and all three models, MoleCode consistently improved optimization quality~(Fig.~\ref{fig:results_2_chemistry}a). The benefit of MoleCode was the most pronounced for Gemini-2.5-Flash, which achieved 0.0 mean LogP improvement with SMILES whereas a +1.15 improvement with MoleCode. Similar increases of mean LogP improvement were seen in Gemini-3-Pro and GPT-5.2, at +0.66 and +0.20 over SMILES, respectively. Solubility optimization also benefited from MoleCode consistently, with corresponding increases of +0.38, +0.21, and +0.34. The distributions of optimization gains shifted toward larger improvements under MoleCode for both objectives~(Fig.~\ref{fig:results_2_chemistry}b), indicating that the effect was broadly distributed rather than driven by a small number of outliers. All references to ``LogP'' in this study denote ``penalized LogP'', as defined in prior works~\cite{kusner2017grammar, jin2018junction} and described in the Methods section.

We then examined whether the edits produced under MoleCode were chemically interpretable. Descriptor-level analysis showed directionally coherent structural changes aligned with the target objective~(Fig.~\ref{fig:results_2_chemistry}c). For LogP optimization, MoleCode preferentially reduced hydrogen-bond donor and acceptor counts while increasing halogenation, changes consistent with increased lipophilicity. For solubility optimization, this pattern shifted in the expected direction, with increased hydrogen-bond donor tendency and reduced halogenation. Structural conservation analysis further showed that MoleCode maintained high Tanimoto similarity to the starting molecules while producing substantial property improvements~(Fig.~\ref{fig:results_2_chemistry}c). This suggests that explicit graph structure supports targeted modification rather than broad unconstrained rewriting.

A representative LogP optimization example illustrates this behaviour~(Fig.~\ref{fig:results_2_chemistry}d). Starting from the same molecule, the SMILES-prompted model produced a structurally degraded candidate with $\Delta\mathrm{LogP}=-0.054$. By contrast, MoleCode identified explicit graph nodes and proposed localized, chemically interpretable edits, including amide nitrogen modification and ether-to-methylene substitutions, yielding a substantially improved molecule with $\Delta\mathrm{LogP}=+2.360$.

These results extend the structural generalization finding from molecular perception to molecular design. By exposing atoms and bonds as editable objects, MoleCode enables LLMs not only to read molecular structure more reliably, but also to modify it in ways that are localized, property-aligned and chemically interpretable.

\begin{figure*}[!th]
\centering
\includegraphics[width=\textwidth]{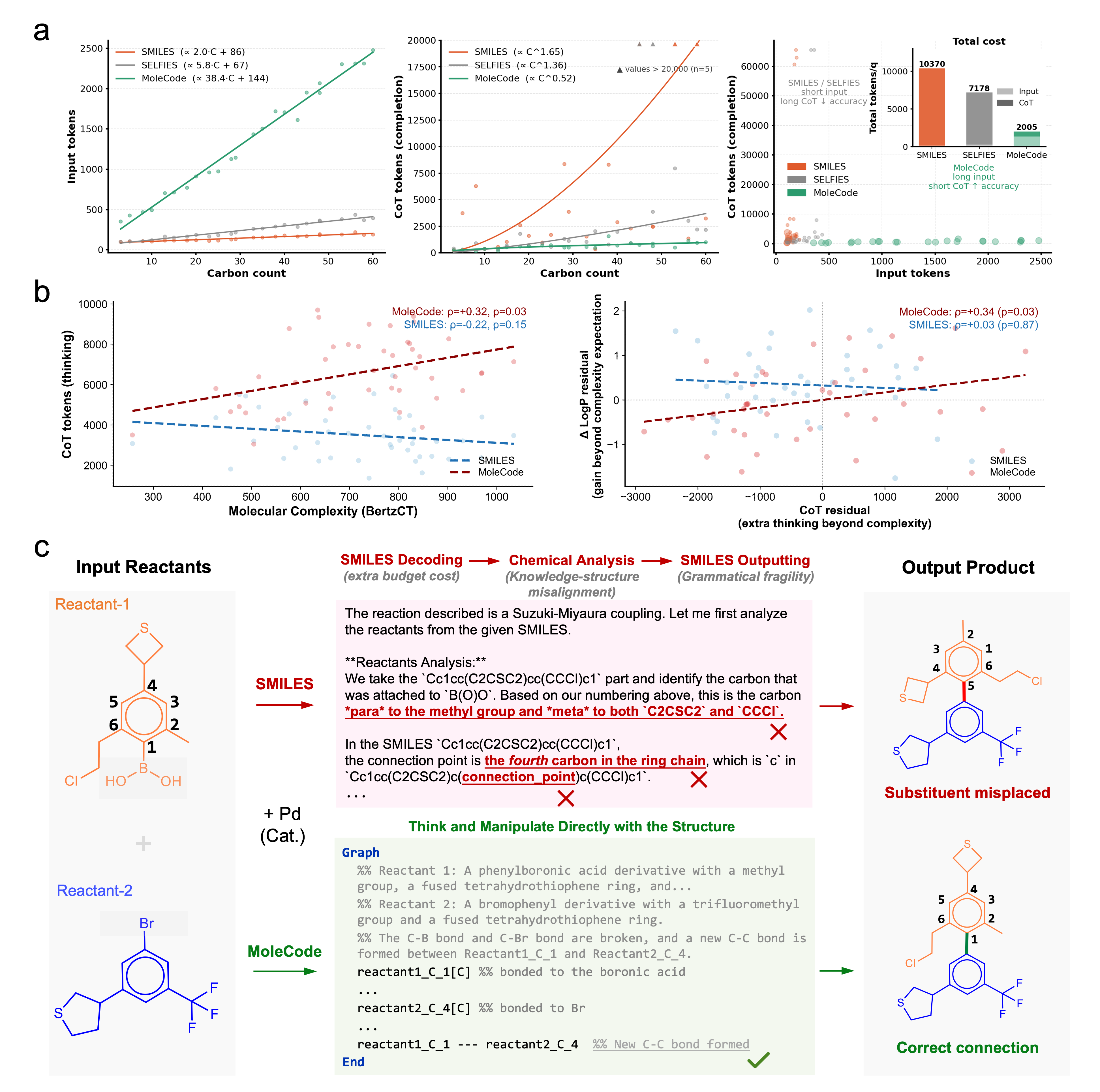}
\caption{
\small
\textbf{MoleCode reallocates inference from structural reconstruction to productive reasoning over explicit structure.}
\textbf{a}, Token-cost decomposition across molecular representations for Gemini-2.5-Flash. MoleCode uses more input tokens because topology is stated explicitly, but its chain-of-thought~(CoT) length grows sub-linearly with molecular size, unlike the super-linear growth observed for SMILES and SELFIES. This shifts inference into a long-input/short-CoT regime with lower total token cost.
\textbf{b}, Productive and unproductive reasoning during molecular optimization. MoleCode elicits longer CoT as molecular complexity increases, and this additional reasoning is positively correlated with optimization gain. In contrast, thinking longer with SMILES does not translate into productivity.
\textbf{c}, Qualitative comparison of reasoning chains for the same molecule under SMILES and MoleCode. SMILES reasoning spends much of its effort recovering implicit connectivity and resolving structural ambiguity, whereas MoleCode reasoning directly uses explicit atom and bond annotations to identify chemically meaningful constraints and make localized edits.
}
\label{fig:results_3_scaling}
\end{figure*}

\subsection*{MoleCode reallocates inference towards productive reasoning}

MoleCode uses longer inputs than conventional molecular strings because topology is written explicitly in the prompt. We therefore asked whether this verbosity increases total inference cost, or whether explicit structure changes how computation is allocated during reasoning.

Input token count scaled approximately linearly with molecular size for all representations, but MoleCode inputs were longer than SMILES and SELFIES because atoms and bonds are represented as graph primitives~(Fig.~\ref{fig:results_3_scaling}a, left). This increased upfront cost was offset by a substantial reduction in chain-of-thought~(CoT)~\cite{wei2022cot} reasoning tokens~(Fig.~\ref{fig:results_3_scaling}a, middle). With SMILES and SELFIES, CoT token cost increased superlinearly with carbon count, at approximately $C^{1.65}$ and $C^{1.36}$, respectively. With MoleCode, CoT token cost scaled sublinearly, approximately at $C^{0.52}$. Therefore, MoleCode shifts inference into a long-input and short-reasoning regime, whereas SMILES and SELFIES occupy a short-input and long-reasoning regime associated with lower accuracy. Despite using the longest prompts, MoleCode achieved the lowest total token cost per query because the reduction in reasoning-token generation largely outweighed the increase in input length~(Fig.~\ref{fig:results_3_scaling}a, right). These findings are consistent with the interpretation that explicit structure externalizes part of the structural reconstruction burden into the prompt itself.

We next asked whether the remaining reasoning effort under MoleCode was more closely associated with successful optimization. During molecular optimization, reasoning length under MoleCode increased with molecular complexity~($\rho=0.32$, $p=0.03$), whereas SMILES showed no meaningful positive relationship~($\rho=-0.22$, $p=0.15$; Fig.~\ref{fig:results_3_scaling}b). More importantly, complexity-adjusted reasoning-length residuals correlated positively with optimization gain under MoleCode~($\rho=0.34$, $p=0.03$), but not under SMILES~($\rho=0.03$, $p=0.87$), as shown in Fig.~\ref{fig:results_3_scaling}b. In summary, longer reasoning with MoleCode leads to better optimization, but this does not apply to SMILES.

A qualitative side-by-side comparison supported this view~(Fig.~\ref{fig:results_3_scaling}c). Under SMILES, the model spent much of its reasoning trace recovering implicit connectivity and resolving structural ambiguity. Under MoleCode, the model directly used atom and bond annotations to identify chemically meaningful constraints, reason about local structural consequences and propose localized edits.

Together, these results suggest a functional interpretation of MoleCode's advantage. Under implicit string representations, a substantial fraction of inference appears to be devoted to reconstructing hidden topology from sequential syntax. Under MoleCode, topology is already accessible in the input, allowing generated reasoning tokens to operate more directly on chemistry-relevant operations. MoleCode therefore does not merely change the format of the molecule; it changes how inference is allocated during molecular reasoning.

\subsection*{MoleCode scales to large and repetitive polymers}

\begin{figure*}[!th]
\centering
\includegraphics[width=\textwidth]{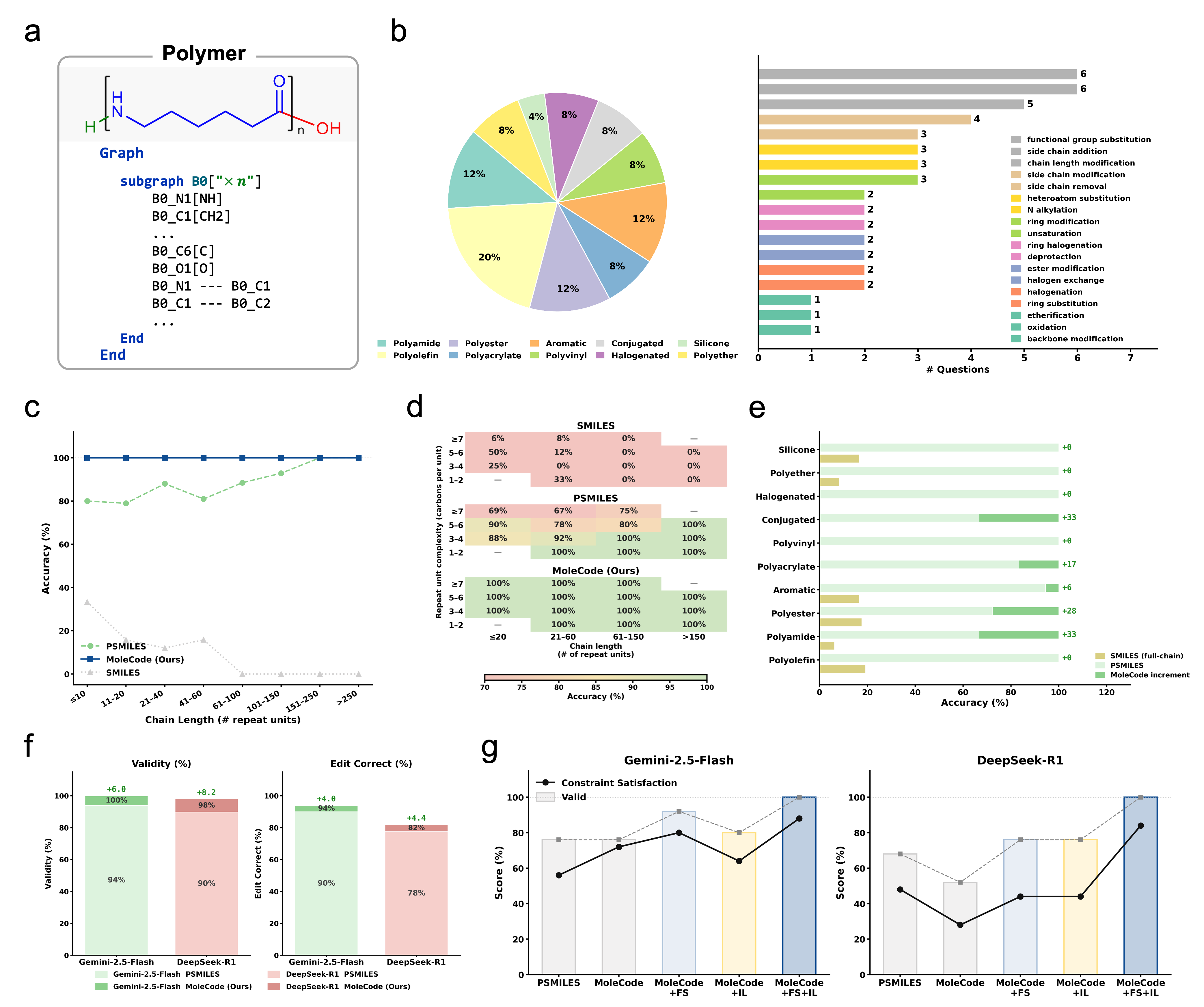}
\caption{
\small
\textbf{The benefit of explicit structure becomes stronger as chemical objects become larger and more repetitive.}
\textbf{a}, MoleCode representation of a polymer repeat unit. A polymer can be written as repeated subgraphs with multiplicity $\times n$, avoiding expansion of the full chain while preserving atom and bond connectivity inside the repeating motif.
\textbf{b}, Overview of the ChemLLM-Polymer benchmark, covering polymer classes and edit-type distributions for three evaluation settings: structure scaling through carbon counting across polymer chain lengths and polymer classes, polymer editing across multiple chemically meaningful edit types, and \textit{de novo} polymer generation under structural constraints.
\textbf{c}, Generalization across polymer chain length (number of repeat units) for Gemini-2.5-Flash. MoleCode maintains near-perfect accuracy across all lengths, while full-chain SMILES collapses to 0\% as the polymer chain becomes longer.
\textbf{d}, Carbon-counting accuracy jointly stratified by polymer chain length and repeat-unit complexity (carbons per repeat unit). Full-chain SMILES degrades along the chain-length axis, PSMILES degrades along the repeat-unit-complexity axis, and MoleCode remains uniformly accurate across both axes, isolating the orthogonal failure modes of the two baselines.
\textbf{e}, Carbon-counting accuracy stratified by polymer class. Across diverse polymer families, MoleCode consistently improves over PSMILES, with the largest gains appearing in structurally repetitive and chemically complex polymer classes.
\textbf{f}, Polymer editing performance under PSMILES and MoleCode representations. MoleCode improves both structural validity and edit correctness.
\textbf{g}, \textit{De novo} polymer generation ablation across prompting and planning strategies. Constraint satisfaction and structural validity are evaluated for PSMILES, MoleCode, MoleCode with few-shot examples (FS), MoleCode with interleaved planning (IL), and MoleCode with both few-shot examples and interleaved planning (FS+IL). Combining MoleCode with structured planning and demonstrations yields the strongest generation performance across both models.
}
\label{fig:results_4_long_molecules}
\end{figure*}

We next investigated whether the advantages of explicit structure extend beyond small molecules to polymers, where long chains and repeated motifs amplify the difficulty of structural reasoning. MoleCode represents polymers by declaring the repeat unit as an explicit subgraph and attaching a multiplicity operator such as $\times n$, rather than expanding every repeat into a long full-chain string. For example, the nylon-like repeat unit in Fig.~\ref{fig:results_4_long_molecules}a is written as a compact graph object whose internal atoms and bonds remain explicit while chain length is controlled symbolically by $n$. To evaluate this setting, we introduced the ChemLLM-Polymer benchmark, comprising three tasks: structure scaling through carbon counting, polymer editing and \textit{de novo} polymer generation~(Fig.~\ref{fig:results_4_long_molecules}b).

The first task evaluated carbon-counting accuracy across polymer chains of increasing length and across ten polymer classes. As polymer size increased, MoleCode substantially outperformed both full-chain SMILES and PSMILES~\cite{PSMILES}, a polymer-specific extension of SMILES designed to represent repeating structures compactly~(Fig.~\ref{fig:results_4_long_molecules}c). MoleCode achieved near-perfect accuracy across all chain-length bins, including polymers exceeding 250 repeat units. Full-chain SMILES collapsed at larger scales, reflecting the burden of very long expanded strings. PSMILES remained more compact, but was less accurate because the model still had to infer repeat-unit chemistry and multiplicity from an implicit sequence.

Joint stratification by chain length and repeat-unit complexity separated two failure modes of implicit polymer strings~(Fig.~\ref{fig:results_4_long_molecules}d). Full-chain SMILES degraded primarily as the number of repeat units increased. PSMILES was more stable along the chain-length axis, but degraded as the repeat unit became more complex. MoleCode remained accurate across both axes because the repeat unit was explicit and the repetition count was represented symbolically.

Stratifying carbon-counting performance by polymer class revealed the same pattern across chemically diverse families~(Fig.~\ref{fig:results_4_long_molecules}e). MoleCode consistently improved over PSMILES across polyamides, polyesters, polyacrylates, silicones, halogenated polymers and conjugated systems. The largest gains appeared in polymer families with highly repetitive or chemically complex structures, including halogenated and aromatic systems. This supports the view that explicit topology becomes increasingly valuable as repetition and compositional complexity increase.

We next evaluated polymer editing tasks spanning eighteen chemically meaningful edit types, including side-chain modification, ring substitution, oxidation and backbone modification~(Fig.~\ref{fig:results_4_long_molecules}f). MoleCode improved both structural validity and edit correctness across Gemini-2.5-Flash and DeepSeek-R1. Thus, explicit graph structure improved not only structural perception, but also manipulation fidelity in macromolecular settings.

Finally, we evaluated \textit{de novo} polymer generation under constraint-based and fragment-assembly settings with Easy, Medium and Hard difficulty levels~(Fig.~\ref{fig:results_4_long_molecules}g). We compared PSMILES, MoleCode, MoleCode with few-shot demonstrations~(FS), MoleCode with interleaved planning~(IL) and the combination of both strategies~(FS+IL). The best performance was achieved when MoleCode was combined with both strategies.

Collectively, these experiments show that the advantage of MoleCode is not limited to small molecules. As chemical objects become larger, more repetitive, and more compositionally structured, implicit sequential encodings degrade increasingly rapidly, whereas explicit graph structure remains stable and operable.

\subsection*{MoleCode extends from molecules to broader scientific structure}

The experiments above establish MoleCode as an effective interface for molecular reasoning. We next asked whether the same graph-native abstraction can extend beyond individual molecules to higher-order forms of chemical structure.

\begin{figure*}[!tb]
\centering
\includegraphics[width=\textwidth]{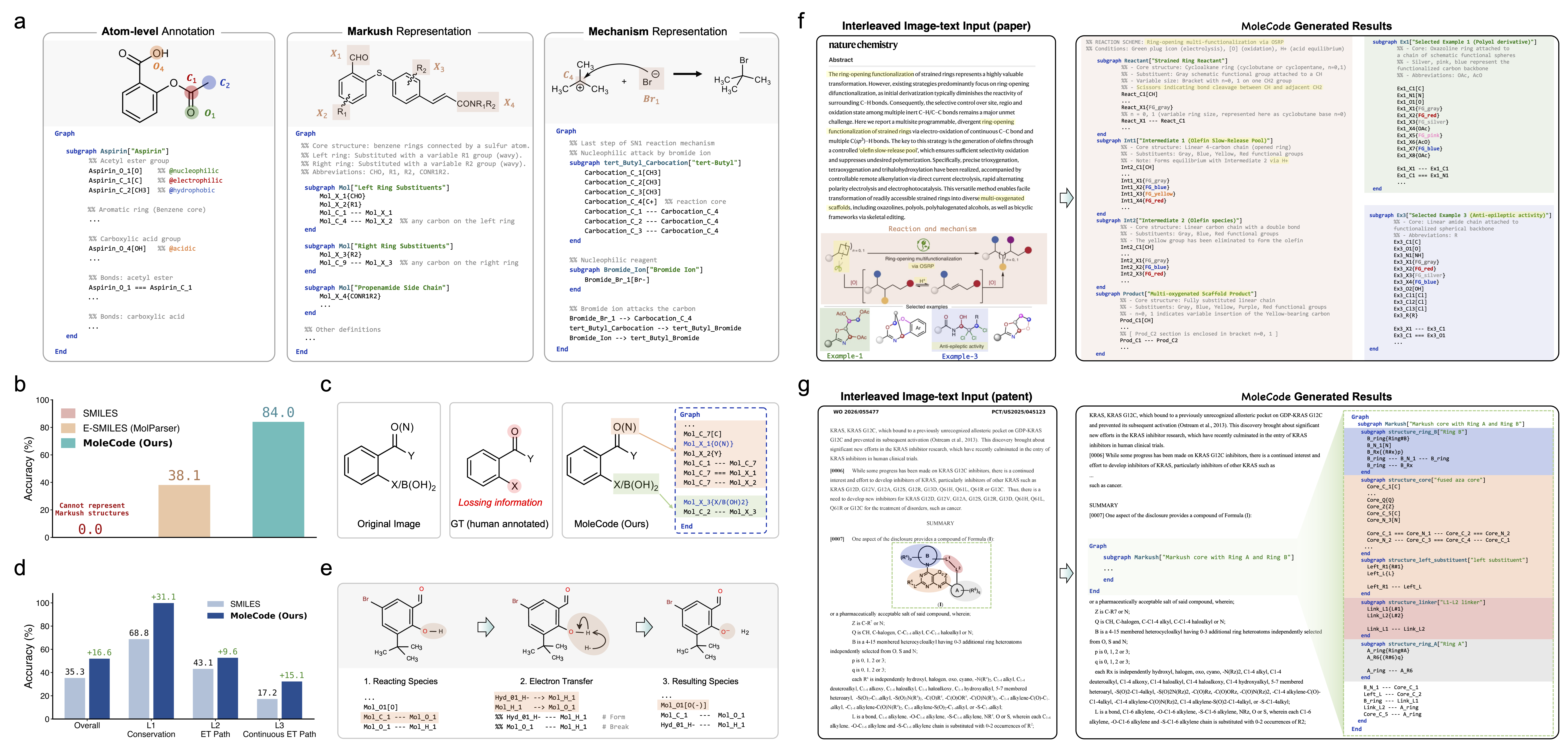}
\caption{
\small
\textbf{MoleCode is a general language for scientific structure.}
\textbf{a}, One grammar spans atom-level molecular annotations, Markush structures with variable substituents and reaction mechanisms with graph transformations.
\textbf{b}, On Markush representation, MoleCode substantially outperforms SMILES, which cannot represent Markush structures, and E-SMILES, which captures them only partially.
\textbf{c}, In a Markush case study, MoleCode preserves substituent logic and structural relations that are lost in simplified human E-SMILES annotations.
\textbf{d}, On mechanism representation, MoleCode improves overall accuracy, conservation, electron-transfer paths and continuous electron-transfer paths.
\textbf{e}, In a mechanism case study, MoleCode represents reacting species, electron transfer and resulting species as explicit, local graph edits.
\textbf{f}, MoleCode parses interleaved image-text content from a recently published chemistry paper~\cite{li2026programmable} into a unified structural graph.
\textbf{g}, MoleCode similarly parses interleaved image-text content from a recent patent disclosure~\cite{wo2026055477} containing Markush cores, substituent definitions and image-text descriptions.
Colored block highlights in \textbf{f} and \textbf{g} were added for visualization and are not present in the original documents.
}
\label{fig:results_5_extension}
\end{figure*}

MoleCode is built from a general Subgraph--Node--Edge abstraction in which graph nodes represent typed entities and edges represent explicit relations. Because this abstraction is not tied to a single molecular scale, the same syntax can represent multiple levels of chemical structure, including atom-level annotations, Markush structures, and mechanism-style graph transformations~(Fig.~\ref{fig:results_5_extension}a).

At the most concrete level, MoleCode supports atom-level annotation in which chemically meaningful attributes such as nucleophilicity, electrophilicity, acidity and hydrophobicity are attached directly to molecular graph nodes. These annotations are machine-readable, atom-resolved and auditable within the same structural representation~(Fig.~\ref{fig:results_5_extension}a, left).

At a higher level of abstraction, MoleCode supports Markush representations in which substituent positions, attachment sites, alternative groups, and logical relations are expressed as explicit graph nodes and edges~(Fig.~\ref{fig:results_5_extension}a, middle). We evaluated this setting using the MolParser benchmark~\cite{fang2025molparser}, which represents Markush structures with E-SMILES~\cite{fang2025molparser}, an extended and dataset-specific version of SMILES. Although E-SMILES is more expressive than standard SMILES, it captures these structures only partially, omitting complex structural logics such as logical OR (``functional group A'' OR ``functional group B''). By contrast, MoleCode preserves variable groups and substituent logic in a graph-native form and leads to substantial performance gain from 38.1\% to 84.0\%~(Fig.~\ref{fig:results_5_extension}b). A qualitative comparison further shows that E-SMILES annotations can collapse variable regions into simplified placeholders, whereas MoleCode preserves attachment relationships and alternative substituent structure through explicit graph annotations~(Fig.~\ref{fig:results_5_extension}c).

MoleCode also extends to reaction mechanisms, where the relevant structure is not a single static molecule but a sequence of graph transformations~(Fig.~\ref{fig:results_5_extension}a, right). In this setting, reacting species, electron-transfer paths, bond changes and resulting species can be represented as explicit node and edge operations. MoleCode improves mechanism representation across overall accuracy, conservation, electron-transfer paths and continuous electron-transfer paths, with absolute improvements of 16.1\%, 31.1\%, 9.6\% and 15.1\% across these levels~(Fig.~\ref{fig:results_5_extension}d). A representative case shows that MoleCode decomposes the mechanism into reacting species, electron transfer and resulting species while preserving the correspondence between molecular diagrams, curved-arrow operations and local graph edits~(Fig.~\ref{fig:results_5_extension}e).

Finally, we examined whether MoleCode could serve as an interface for multimodal chemistry documents in which chemical information is distributed across text, molecular diagrams and reaction schemes. As a generalization test, we used a frontier multimodal LLM, Gemini-3-Pro, to parse a recent article on programmable divergent electrochemical ring-opening multifunctionalization of strained rings published in Nature Chemistry~\cite{li2026programmable} into a single MoleCode graph~(Fig.~\ref{fig:results_5_extension}f). The article was published in April 2026 and reports a newly developed synthetic strategy, making it unlikely to have been memorized during LLM pretraining, and therefore ensures the task highlights extraction, structural grounding and organization from the supplied document. MoleCode preserved molecular entities, annotations, intermediates, and cross-modal relations within one graph-native representation. Because atoms in molecular images correspond naturally to nodes and bonds correspond to edges, visual chemical structure can be mapped into MoleCode without first reducing it to a purely sequential string.

We further tested the same multimodal parsing ability on a recent patent disclosure containing Markush cores, substituent definitions and image-text descriptions, published in March 2026~\cite{wo2026055477}. MoleCode converts the patent content into a structured graph that preserves scaffold definitions, variable groups, substituent logic and relations between the text and molecular image~(Fig.~\ref{fig:results_5_extension}g). This example suggests that the same representation can organize chemical information not only from research articles, but also from complex disclosure formats where molecular structure is distributed across figures, claims and descriptions.

These examples suggest that MoleCode is not only a molecular string alternative, but also a graph-native interface for higher-order chemical structure. The same abstraction can express molecules, generalized scaffolds, reaction transformations and multimodal chemistry documents within one unified structural language.

\section*{Discussion}

Our results show that molecular reasoning in frontier LLMs is strongly shaped by whether molecular topology is directly accessible within the language input. SMILES is compact and widely adopted, but it hides graph structure behind sequential syntax. MoleCode instead exposes molecular components and relations as explicit linguistic objects, allowing models to operate on topology rather than reconstruct it as an intermediate step. Across small molecules, goal-directed optimization and polymers, the advantage of MoleCode was largest when structural access was most limiting, including unfamiliar compounds, topology-sensitive tasks and long repetitive chemical objects.

This finding reframes the role of representation in language-based molecular science. The contrast is not simply between strings and graphs, or between LLMs and graph neural networks. Graph neural networks provide direct access to topology, but they are usually specialized predictors. LLMs provide flexible reasoning, natural-language interaction and iterative editing, but struggle when structure is hidden inside implicit encodings. MoleCode brings these capabilities closer together by placing graph structure inside the context window as readable and editable language. The model is therefore not asked to infer the molecular graph before reasoning; it is asked to reason over a graph that is already explicitly represented.

The inference analyses further suggest that MoleCode changes how computation is used. Although MoleCode prompts are longer than SMILES, reasoning-token growth scales more efficiently with molecular size, and additional reasoning effort is more closely associated with optimization success. These observations are consistent with the interpretation that explicit topology reallocates inference away from structural reconstruction and towards chemically meaningful operations, such as identifying functional motifs, evaluating local consequences and making targeted edits. Qualitative reasoning traces support this view: SMILES-based reasoning often spends substantial effort resolving implicit connectivity, whereas MoleCode reasoning proceeds more directly from atom and bond annotations to localized chemical decisions.

The same perspective also explains why MoleCode can operate beyond individual molecules. Its Subgraph--Node--Edge abstraction can express atom-level annotations, Markush structures, reaction transformations and multimodal chemistry documents within one graph-native syntax. More broadly, MoleCode illustrates a representational principle for scientific language interfaces. Many scientific objects, including molecules, polymers, reaction schemes, knowledge graphs, circuits and experimental workflows, are fundamentally relational. Compressing them into implicit sequences forces models to recover structure before using it. Exposing structure directly may allow LLMs to apply their reasoning capabilities more reliably across structured scientific domains.

Because MoleCode exposes editable graph objects directly in language, it also supports agentic scientific workflows. As a step towards this direction, we incorporated MoleCode as a coding-agent skill that allows agents to parse molecular inputs, manipulate graph objects, validate edits and convert structures back into standard chemical formats across iterative workflows. We implemented this capability in \textbf{AtomFlow}, a MoleCode-driven human-AI interaction system for accessing, selecting and editing molecular graph objects (please visit our website: \href{https://atomflow-ai.com}{\url{https://atomflow-ai.com}}). The two core applications, \textbf{AtomChat} supports natural-language editing of atoms, bonds and fragments, whereas \textbf{AtomRetro} supports retrosynthesis planning over explicit MoleCode structures. These applications illustrate how graph-explicit molecular language can move LLM-based chemistry from molecule-level prompting towards atom-level interaction, where reasoning, editing and validation operate on the same structured object.

Several limitations remain. MoleCode does not create chemical knowledge that a model lacks. Smaller or less chemically capable models may still generate invalid structures or violate chemical constraints, indicating that explicit representation must be paired with sufficient domain knowledge. MoleCode is also more verbose than SMILES, and specialized domains such as organometallic chemistry, inorganic solids and biological macromolecules may require additional primitives for coordination geometry, periodicity and higher-order organization. These limitations suggest a natural next step: incorporating graph-explicit representations into molecular and scientific pretraining, so that future models learn structured domains through explicit topology from the outset rather than through implicit string reconstruction.

Taken together, these findings suggest that representation design is not a peripheral engineering choice for scientific LLMs. It determines which aspects of a structured object are immediately available for reasoning, which operations are local and auditable, and how much computation is spent recovering structure rather than reasoning over it. MoleCode demonstrates this principle in molecular science, suggesting that future scientific language interfaces may benefit from representing structured objects in forms that are directly operable rather than implicitly encoded.

\section*{Methods}

\subsection*{MoleCode grammar for small molecule, polymer, Markush, and mechanism}

MoleCode uses one unified grammar across all use cases: the core grammar and conversion examples in Fig.~\ref{fig:concept}b-d, the polymer repeat-unit representation in Fig.~\ref{fig:results_4_long_molecules}a, and the higher-order structural extensions in Fig.~\ref{fig:results_5_extension}a. MoleCode is implemented as a text serialization of molecular graphs using three primitives: \texttt{Subgraph}, \texttt{Node} and \texttt{Edge}. A subgraph defines a scoped structural object, such as a molecule, repeat unit, Markush scaffold, variable substituent or reaction intermediate. A node represents either an atom or a higher-level labelled chemical entity and carries a persistent identifier that remains stable across prompt construction, model reasoning and post-processing. An edge represents an explicit relation between two nodes. For molecules, edges correspond to bonds and encode bond order directly; the implementation uses Mermaid-compatible operators, including \codeinline{---} for single bonds, \codeinline{===} for double bonds and \codeinline{-.-} for triple bonds. Aromatic small-molecule inputs are generally serialized in Kekul\'{e} form, whereas polymer editing and generation prompts use \codeinline{<-->} to mark aromatic bonds when this improves syntactic clarity. Inline comments beginning with \codeinline{\%\%} are permitted inside MoleCode blocks and are ignored by the parser; these comments allow models to attach reasoning, atom lists and validation checks to the graph without changing the parsed structure.

Conversion between standard molecular formats (such as SMILES and the RDKit MOL file) and MoleCode is deterministic and does not use a learned model. For small molecules, input SMILES are parsed and sanitized with RDKit~\cite{rdkit}, stereochemical information is assigned when present, aromatic systems are Kekul\'{e}ized according to the representation mode, and every heavy atom is assigned a namespace-local identifier of the form \codeinline{Molecule\_C\_4} or \codeinline{Reactant1\_N\_2}. Node labels encode element identity and, when used by the task, hydrogen count, formal charge and stereochemical suffixes. Bond lines are emitted from the RDKit bond table with explicit source and target identifiers.

For Markush and document-level examples, MoleCode additionally allows abbreviation nodes written with labelled braces, such as \codeinline{R1}, \codeinline{Ar}, \codeinline{Boc} or \codeinline{(CH2)n}. These nodes preserve visible chemical labels that cannot be represented faithfully by ordinary SMILES. So we refer to E-SMILES~\cite{fang2025molparser} for implementation. Variable substituents, attachment sites, logical alternatives and non-expanded groups are encoded as explicit nodes and typed edges rather than collapsed into an anonymous wildcard. For polymers, repeat units are represented as subgraphs carrying a multiplicity label such as \colorbox{gray!10}{\texttt{$\times n$}}, together with two terminus markers corresponding to the two \codeinline{*} atoms in PSMILES~\cite{PSMILES}. Hydrogen counts for polymer editing and generation are inferred during reverse conversion from valence and connectivity rather than read directly from node labels, which reduces failures caused by minor hydrogen-bookkeeping drift.

Reaction mechanisms are represented in MoleCode as graph-valued reaction paths. Each mechanism is encoded as an ordered sequence of chemically meaningful states and transformations: a state contains the molecular species present at a given point along the reaction coordinate, whereas a transformation describes the structural and electronic changes connecting two adjacent states. The representation preserves key mechanistic objects, including reactants, products, intermediates and transition states, while maintaining atom-level identity throughout the full path. Electron-transfer relations between atoms, bonds or lone-pairs are encoded explicitly, enabling direct representation of bond formation and cleavage, bond-order changes, charge migration, proton transfer, resonance and coordination events. Generic substituents and unresolved groups are retained as labelled nodes when their internal structures are not involved in the mechanism.

\subsection*{Molecule-related benchmark}
The molecule-related benchmark underlying Fig.~\ref{fig:results_1_main}a is organized into four task families: molecular editing, molecular generation, molecular understanding and molecular analysis. Molecular editing tasks, including atom addition, deletion and substitution, derive from the TOMG benchmark~\cite{li2024speak}. Molecular generation tasks include atom-number-constrained generation and functional-group-guided generation from TOMG, together with text-based \textit{de novo} molecule generation from ChEBI-20~\cite{molt5}. Molecular understanding and analysis tasks are drawn from subsets of ChemIQ~\cite{ChemIQ} and ChemCoTBench~\cite{ChemCoTBench}; these tasks include carbon counting, molecular formula prediction, IUPAC-to-SMILES conversion, reaction prediction and NMR elucidation. We additionally use molecular equivalence, functional-group counting, ring counting and ring-system scaffold identification to evaluate representation variants in the ablation analysis.

For tasks derived from ChemIQ~\cite{ChemIQ} and ChemCoTBench~\cite{ChemCoTBench}, the source files contain 806 questions across eight task labels. The experiments use the subsets relevant to the main figure: carbon counting, ring counting, reaction prediction, NMR elucidation, and SMILES/IUPAC conversion. Carbon and ring counting are scored by exact integer match. Molecular formula prediction is scored by exact formula match after normalization. IUPAC-to-SMILES, reaction prediction and NMR elucidation outputs are canonicalized with RDKit and scored by canonical-SMILES equality to the reference molecule or product. \textit{Outputs that fail extraction or RDKit parsing are counted as incorrect.}

For TOMG and ChEBI-20 generation or editing tasks, each natural-language instruction is paired with the same source molecule or generation constraint across representations. SMILES and SELFIES inputs are provided as linear strings. MoleCode inputs are generated from the RDKit molecular graph and expose atom identifiers, atom labels and bond relations directly in the prompt. For editing tasks, outputs are converted to canonical SMILES and evaluated for validity and satisfaction of the requested structural edit. For generation tasks, model outputs are parsed from the requested representation and converted back to canonical SMILES when a structural answer is required. For tasks requiring a natural-language or numeric answer, the leading answer span is extracted using task-specific regular expressions and compared with the reference answer.

\subsection*{Familiarity, complexity and representation ablations}

For the familiarity, complexity and representation-ablation analyses in Fig.~\ref{fig:results_1_main}b-d, we test whether MoleCode improves structural generalization rather than memorization of familiar molecular strings using a PubChem-based memorization benchmark that stratifies molecules into three exposure tiers (Fig.~\ref{fig:results_1_main}b). The tiers are defined by PubChem Compound Identifier~(CID), which serves as a proxy for deposition time and downstream web or literature presence~\cite{kim2023pubchem}: popular/famous compounds use CIDs 1--10,000, classical/medium-exposure compounds use CIDs 10,001--200,000, and novel/obscure compounds use CIDs 130,000,000--165,000,000. Candidate metadata are retrieved with PubChem PUG-REST batch property queries for molecular formula, molecular weight, canonical or connectivity SMILES and IUPAC name. Famous-tier CIDs are enumerated sequentially; medium-tier candidates are drawn as a uniform random sample of 8,000 CIDs; obscure-tier candidates are first subsampled at stride 100 across the target CID interval and then uniformly sampled to 10,000 CIDs. Retrieval uses batches of 200 CIDs with a 0.25--0.30-s inter-request delay. Molecules are retained only if RDKit parses the SMILES, the molecule is a single connected component, all atoms belong to the organic subset \{C, H, N, O, S, F, Cl, Br, P, I\}, at least one carbon atom and at least three heavy atoms are present, and PubChem-reported molecular weight lies in the range 50--500 Da. To control for molecular size, candidates are stratified into four molecular-weight bins, [50,150), [150,250), [250,350) and [350,500) Da. Within each tier, 13 molecules are sampled per bin and then uniformly trimmed to exactly 50 molecules per tier, yielding 150 molecules with approximately matched molecular-weight distributions. All sampling uses random seed 42. For each retained molecule we generate RDKit-canonical SMILES, three random but equivalent SMILES strings, canonical MoleCode, atom-permuted MoleCode and the Hill-order molecular formula used as the ground-truth answer. Continuous familiarity covariates, including PubChem synonym count and PubMed cross-reference count, are also collected and used on a log scale when regression-style analyses are required. Molecular complexity is analysed using molecular weight, carbon count and, for optimization-token analyses, BertzCT graph complexity.

Representation ablations compare SMILES with progressively richer MoleCode variants on molecular understanding tasks using the same questions, models and answer extractors for all variants. The base MoleCode variant encodes atom and bond declarations only. The ring-aware variant adds explicit ring-level annotations so that ring membership and ring closures are available without reconstructing them from edge patterns. The scaffold-aware variant further exposes ring-system scaffolds, scaffold membership and substituent attachment relationships as graph-level annotations. These variants are evaluated on molecular equivalence, functional-group counting, ring counting and ring-system scaffold identification with DeepSeek-R1 and Gemini-2.5-Flash. Molecular equivalence and ring-system scaffold identification are reported as accuracy; functional-group counting and ring counting are reported as mean absolute error. Numeric answers are extracted directly and compared with the reference counts, whereas structural answers are canonicalized before equivalence scoring.

Scaling with molecular size is analysed in the carbon-counting task. For the expanded scaling experiment, 200 examples spanning carbon counts from $C=3$ to $C=102$ are sorted by expected carbon count and divided into 20 bins of 10 examples each. Bucket-wise accuracy is plotted after LOWESS smoothing with fraction 0.35 for each representation and model. The three molecular-size regimes used for interpretation are $C=3$--27, $C=28$--52 and $C=53$--102. The same 40-example subset, sampled at regular intervals from the carbon-counting set and covering $C=3$--60, is used for token-cost analysis with Gemini-2.5-Flash.

\subsection*{Molecular optimization benchmark}

The molecular-optimization experiments shown in Fig.~\ref{fig:results_2_chemistry}a-d use the ChemCoTBench~\cite{ChemCoTBench} molecular-optimization benchmark. For each property, 100 source molecules are evaluated, with paired SMILES and MoleCode prompts generated from the same molecule. The benchmark includes LogP, QED, and ESOL-style aqueous solubility in the raw experiments; the main figure reports LogP (normalized penalized version~\cite{kusner2017grammar, jin2018junction}) and solubility, using the benchmark's provided metrics. Penalized LogP is defined as $\text{score}(m) = \log\text{P}(m) - \text{SA}(m) - \text{cycle}(m)$, which subtracts the synthetic accessibility~(SA) score and number of long cycles from the octanol-water partition coefficients~\cite{jin2018junction}. Each prompt describes the property objective, provides the source molecule in the corresponding representation and asks the model to propose an improved target molecule. SMILES-mode responses are parsed from a JSON field named \codeinline{Final Target Molecule}; MoleCode-mode responses are parsed as graph blocks and converted back to RDKit molecules before scoring.

Optimization experiments run on Gemini-2.5-Flash, Gemini-3-Pro and GPT-5.2 with a maximum token budget of 90,000, extended to 200,000 for long constrained-design case studies. Each source molecule is evaluated with a single successful call per representation and model. Property improvement is computed as $\Delta = s(\mathrm{target}) - s(\mathrm{source})$, where $s$ is the property-specific scoring function. A target is scored only if it can be parsed by RDKit; invalid outputs contribute $\Delta=0$ to aggregate bar plots. We also record validity and success rates, where success requires a valid target with improved property value relative to the source.

Distributional analyses use the same per-molecule $\Delta$ values. For each model, property and representation, histograms are computed using 16 density-normalized bins and visualized as Gaussian kernel-density estimates with bandwidth 0.4, using \codeinline{scipy.stats.gaussian\_kde}. Descriptor-level mechanism analysis is performed on valid source-target pairs with RDKit. We compute changes in hydrogen-bond donor count, hydrogen-bond acceptor count and halogen count, and structural conservation is measured by Tanimoto similarity between Morgan fingerprints with radius 2 and 2048 bits.

\subsection*{Inference-token and reasoning-productivity analyses}

To quantify the inference-allocation patterns shown in Fig.~\ref{fig:results_3_scaling}a-c, token-cost analyses measure how much computation is allocated to the input prompt versus model-generated reasoning and answer tokens. For the carbon-counting analysis, the same 40 examples are run with Gemini-2.5-Flash under SMILES, SELFIES and MoleCode. The token budget is set to 65,500 to avoid truncating long reasoning traces. We record prompt tokens, completion tokens and correctness for each call. When the API exposes hidden reasoning tokens only as usage counts, chain-of-thought token count is estimated as completion tokens minus visible answer tokens. Input-token scaling is fitted as a linear function of carbon count, and reasoning-token scaling is fitted by a power law $y=aC^b$ on log-transformed counts. These fits produce the exponents reported in the main text for SMILES, SELFIES and MoleCode.

For molecular optimization, token-productivity analysis uses a paired 50-molecule LogP subset evaluated under SMILES and MoleCode with Gemini-2.5-Flash. Molecular complexity is quantified by BertzCT. For the relationship between reasoning length and molecular complexity, CoT token counts are plotted against BertzCT and outliers with absolute z-score greater than 2.0 are removed. Linear fits are used for visualization, and Spearman rank correlation is used for inference. To test whether extra reasoning is productive rather than merely associated with harder molecules, both CoT token count and $\Delta\mathrm{LogP}$ are residualized against BertzCT using a linear fit. Spearman correlation is then computed between the two residuals on the paired-valid subset; outputs invalid in either representation are excluded from this paired residual analysis. Per-mode outliers with absolute z-score greater than 1.75 on either residual axis are removed before reporting $\rho$ and $p$ values.

\subsection*{Polymer benchmark}

For the polymer analyses summarized in Fig.~\ref{fig:results_4_long_molecules}a-g, we use a ChemLLM-Polymer benchmark constructed from 25 canonical polymers spanning chemically diverse classes, including polyolefins, halogenated polymers, polyethers, polyvinyls, polyacrylates, polyesters, polyamides, aromatic and conjugated polymers, and silicones. For every question, paired inputs are generated in PSMILES, full-chain SMILES when applicable, and MoleCode, so that the molecular object and instruction are held fixed while the representation changes.

The benchmark comprises three tasks. The carbon-counting task contains 150 examples: each of the 25 polymers is instantiated at six chain lengths chosen so that the total carbon count targets approximately 50, 100, 200, 350, 550 and 800 carbon atoms. Full-chain SMILES baselines are generated by explicit RDKit concatenation of $n$ repeat units after removing attachment-point atoms; termini are excluded from the ground-truth carbon count. The editing task contains 50 natural-language editing instructions spanning side-chain modification, halogenation, functional-group substitution, heteroatom substitution, ring modification, backbone modification, oxidation, deprotection and related categories. The \textit{de novo} generation task contains 25 prompts: 12 constraint-satisfaction tasks and 13 fragment-assembly tasks, each tagged as Easy, Medium or Hard.

PSMILES-to-MoleCode conversion is implemented by parsing the repeat-unit SMILES with RDKit, identifying the two wildcard attachment-point atoms, assigning block-local identifiers to non-wildcard heavy atoms and writing the repeat unit as a subgraph between terminus markers \codeinline{TL} and \codeinline{TR}. Reverse conversion parses node and edge lines, reintroduces two wildcard atoms, constructs an RDKit molecule and canonicalizes it to PSMILES. The converter is validated by round-tripping all 50 editing inputs and requiring equality after canonicalization. Counting prompts are scored by exact integer match. Editing and generation outputs are valid only if they parse, contain exactly two wildcard attachment points and satisfy all required SMARTS checks while matching none of the prohibited SMARTS checks. Canonical PSMILES equality to the reference answer is also recorded as a stricter secondary metric.

Polymer experiments evaluate Gemini-2.5-Flash and DeepSeek-R1. Decoding uses \codeinline{temperature}=0. The maximum token budget is 65,536 for counting and 16,384 or 32,768 for editing and generation, depending on the model. Calls run in parallel with five workers, and partial results stream to disk with per-example identifiers so interrupted runs can be resumed. Polymer generation compares PSMILES, MoleCode, MoleCode with two few-shot examples, MoleCode with interleaved planning comments and MoleCode with both few-shot examples and interleaved planning. Interleaved plans are written as \codeinline{\%\%} comments listing intended atoms, bond counts and valence checks, so they influence generation while remaining invisible to the graph parser.

\subsection*{Markush structures, mechanism and multimodal chemistry documents}

The higher-order structure experiments in Fig.~\ref{fig:results_5_extension}a-g include Markush recognition, mechanism parsing and multimodal document parsing (for both academic papers and patents); for the Markush component, we use the Markush subset of WildMol-10k released with the MolParser-7M benchmark~\cite{fang2025molparser}. We sample 100 examples deterministically with random seed 42 from the \codeinline{test\_markush\_10k} split. The input to the model is a chemical-structure image, and the output is a machine-readable representation. Three formats are compared: plain SMILES, E-SMILES generated by the specialized MolParser-Base system, and MoleCode generated by a general-purpose multimodal model prompted with the MoleCode specification. Plain SMILES is treated as a lower-bound format because it cannot preserve labelled R groups and other Markush variables. E-SMILES is treated as the domain-specific baseline because it is the native annotation format of the benchmark.

Automatic Markush evaluation converts both predictions and E-SMILES ground truth into a common graph representation and applies VF2-style graph isomorphism with custom node and edge matching. Atom labels are compared after normalizing hydrogen counts and charges. Abbreviation labels are normalized by case-folding and by stripping bracket or index decoration so that, for example, \codeinline{R[1]} and \codeinline{R1} can match. Bond order is required to match; stereochemistry is ignored for the main Markush score because many generic structures do not specify wedges consistently. When a predicted abbreviation does not directly match the ground truth, a curated expansion table is used to test whether the expanded graph is isomorphic. Final reported Markush accuracy uses expert manual verification: a chemistry expert reviews the original image, ground truth and prediction for every example and labels each output as correct, partial or wrong. The expert-reviewed correct rate is used in the main figure.

We construct the mechanism-level benchmark from reactions provided in MechFinder\cite{chen2024large}. Starting from atom-mapped complete reactions, we apply the reaction template library in MechFinder\cite{chen2024large} to match each overall reaction to a known mechanistic class and decompose it into a sequence of elementary steps, including the corresponding reaction state and step-wise electron-transfer trajectory. We then perform stratified sampling by reaction type and mechanism length to ensure coverage across both chemical classes and reasoning complexity. From the resulting candidate set, we select two examples per reaction category and further restrict the benchmark to mechanisms with no more than 10 steps. This yields a final benchmark of 100 reactions. 
For each benchmark reaction, we provide the model with the full atom-mapped reaction encoded either as MoleCode or as mapped SMILES, and ask it to generate the atom-mapped sequence of electron-transfer events and reaction states. The predicted mechanism is then evaluated along three dimensions: conservation validity, which measures whether the generated intermediates preserve atomic composition and formal charge; electron-transfer path coverage, which measures how many ground-truth electron-transfer events are correctly recovered under an order-preserving matching criterion; and overall mechanistic completeness, which measures whether the electron-transfer sequence is continuously correct from the beginning of the mechanism. Across these metrics, we find that mechanism parsing based on MoleCode consistently outperforms SMILES-based parsing, showing higher chemical conservation, more accurate electron-transfer prediction, and better recovery of complete reaction mechanisms.

For the multimodal chemistry-document demonstration, we consider a recent chemistry article in which chemical knowledge is distributed across interleaved text, molecular structures and reaction schemes. This setting normally requires manual checking, because the textual descriptions and image-derived molecular structures are cross-modal and are not explicitly aligned with one another. Using Gemini-3-Pro with the MoleCode specification, we show that such mixed chemistry documents can be parsed into a single MoleCode graph that preserves molecular entities, labels, reaction intermediates, image-derived structures and cross-modal relations in one graph-native representation. Human expert verification confirms that the parsed graph accurately recovers the chemical structures in the document.

\subsection*{AtomFlow: MoleCode-driven human-AI interaction agentic system}

We further built MoleCode-powered molecular interaction systems in AtomFlow using an agentic system. Since MoleCode represents every atom, bond and molecular fragment as an explicit and addressable graph object, it can be naturally integrated with the currently powerful coding agents, iteratively refined through agent workflows and converted into validated molecular operations. This makes it possible to move beyond molecule-level prompting towards atom-level human-AI interaction, where users and agents can jointly access, select, operate on and edit the underlying molecular graph. 
Based on this capability, we developed a series of AtomFlow products. \textbf{AtomChat} is a chat-with-molecule system where users can select specific atoms, bonds, or fragments to ask questions or perform arbitrary edits via natural language. Every modification and LLM annotation is reflected in the molecular graph in real-time.
\textbf{AtomRetro} is a retrosynthesis-planning system powered by pure LLM reasoning and the LLM-native molecular language MoleCode, enabling workflows that operate directly over explicit molecular structures. This system accommodates arbitrary chemist preferences, allowing for atom-level comments on synthetic routes while intuitively explaining the rationale and logic for each retrosynthetic step.
%Implementation details of the public interface, including frontend/backend versions and deployment URL, are \textbf{[TO FILL: interactive system details]}.

% \subsection*{Data availability}

% All benchmark datasets, evaluation scripts, and the MoleCode conversion library are available at \url{https://github.com/???}. The interactive demonstration system is available at \url{https://???}.

% \subsection*{Code availability}

% Source code for MoleCode conversion~(\codeinline{rdkit\_to\_mermaid.py}, \codeinline{mermaid\_to\_rdkit.py}), benchmark evaluation~(\codeinline{chemiq\_run.py}), and the interactive interface are available under the MIT license at \url{https://github.com/???}.

\section*{Acknowledgements}
The authors would like to thank Peking University Shenzhen Graduate School for providing the research environment and for its generous support of student entrepreneurship. We are also grateful to AtomFlow Co., Ltd. for the funding and essential support provided for this study. A related patent filing by AtomFlow Co., Ltd. was made at an early stage of this project.

\section*{Author contributions}

Z.Y., C.L. and B.Z. conceived the project. Z.Y. and B.Z. designed the MoleCode syntax and grammar and implemented the conversion algorithms. Z.Y. and C.L. designed the experiments and analyses. C.L., Z.Y. and F.M. organized and framed the paper. Z.Y. conducted and implemented the molecule-related benchmarks, polymer benchmarks and associated analyses. K.L., Z.Y. and B.Z. implemented the document-parsing experiments. Z.Y. and B.Z. implemented the Markush-structure experiments, and Y.W. implemented the mechanism-related experiments. J.Z. implemented the Python version of MoleCode and integrated it into coding agents, and B.Z wrote the instructional skill that teaches coding agents to use MoleCode. H.L. provided analysis insights for molecular optimization and a helpful discussion of computational efficiency. L.L. provided insightful comments on the molecule-related benchmarks and a helpful discussion of the polymer. Z.Y. and C.L. performed the scaling analyses. L.Y. and S.Z. supervised the computational aspects, and F.M. led and supervised the scientific aspects. C.L., Z.Y., B.Z. and Y.W. polished the figures for visualization and demonstration. All authors contributed to writing the manuscript.

% \section*{Competing interests}

% The authors declare no competing interests.

% \section*{Data Availability}

% Data are freely available at \url{}.

% \section*{Code Availability}

% Source code are available under an open-source license at \url{}.

% \bibliographystyle{naturemag}
\bibliographystyle{unsrt}
\bibliography{references}

%%%%%%%%%%%%%%%%%%%%%%%%%%%%%%%%%%%%%
%%%%%% Supplementary Materials %%%%%%
%%%%%%%%%%%%%%%%%%%%%%%%%%%%%%%%%%%%%

% Rename supplementary figures.
\renewcommand{\thefigure}{S\arabic{figure}}
\renewcommand{\theHfigure}{S\arabic{figure}}
\setcounter{figure}{0}
\renewcommand{\thetable}{S\arabic{table}}
\renewcommand{\theHtable}{S\arabic{table}}
\setcounter{table}{0}

\clearpage
\newpage

% \renewcommand\appendixpagename{Supplementary Materials}
% \begin{appendices}

% \begin{figure*}[!ht]
% \centering
% % \includegraphics[width=\textwidth]{figures/???.png}
% \caption{
% Agentic long task.
% }
% \label{fig:supp_agentic}
% \end{figure*}

% \end{appendices}

%% file: references.bib
@article{ashyrmamatov2025survey,
  title={A survey on large language models in biology and chemistry},
  author={Ashyrmamatov, Islambek and Gwak, Su Ji and Jin, Su-Young and Jun, Ikhyeong and Ucak, Umit V and Lee, Jay-Yoon and Lee, Juyong},
  journal={Experimental \& Molecular Medicine},
  pages={1--11},
  year={2025},
  publisher={Nature Publishing Group UK London}
}

@article{bhattacharya2024large,
  title={Large language models as molecular design engines},
  author={Bhattacharya, Debjyoti and Cassady, Harrison J and Hickner, Michael A and Reinhart, Wesley F},
  journal={Journal of Chemical Information and Modeling},
  volume={64},
  number={18},
  pages={7086--7096},
  year={2024},
  publisher={ACS Publications}
}

@article{park2024llamo,
  title={Llamo: Large language model-based molecular graph assistant},
  author={Park, Jinyoung and Bae, Minseong and Ko, Dohwan and Kim, Hyunwoo J},
  journal={Advances in Neural Information Processing Systems},
  volume={37},
  pages={131972--132000},
  year={2024}
}

@article{givechian2026immunostruct,
  title={ImmunoStruct enables multimodal deep learning for immunogenicity prediction},
  author={Givechian, Kevin Bijan and Rocha, Jo{\~a}o Felipe and Liu, Chen and Yang, Edward and Tyagi, Sidharth and Greene, Kerrie and Ying, Rex and Caron, Etienne and Iwasaki, Akiko and Krishnaswamy, Smita},
  journal={Nature Machine Intelligence},
  volume={8},
  pages={70--83},
  year={2026},
  publisher={Nature Publishing Group UK London}
}

@article{zhang1994representations,
  title={Representations in distributed cognitive tasks},
  author={Zhang, Jiaje and Norman, Donald A},
  journal={Cognitive science},
  volume={18},
  number={1},
  pages={87--122},
  year={1994},
  publisher={Wiley Online Library}
}

@article{li2024speak,
  title={Speak-to-Structure: Evaluating LLMs in Open-domain Natural Language-Driven Molecule Generation},
  author={Li, Jiatong and Li, Junxian and Wang, Weida and Liu, Yunqing and Zheng, Changmeng and Zhou, Dongzhan and Wei, Xiao-yong and Li, Qing},
  journal={arXiv preprint arXiv:2412.14642},
  year={2024}
}

@article{zhang1997nature,
  title={The nature of external representations in problem solving},
  author={Zhang, Jiajie},
  journal={Cognitive science},
  volume={21},
  number={2},
  pages={179--217},
  year={1997},
  publisher={Elsevier}
}

@book{olmsted1997chemistry,
  title={Chemistry: the molecular science},
  author={Olmsted, John and Williams, Gregory M},
  year={1997},
  publisher={Jones \& Bartlett Learning}
}

@article{talanquer2022complexity,
  title={The complexity of reasoning about and with chemical representations},
  author={Talanquer, Vicente},
  journal={Jacs Au},
  volume={2},
  number={12},
  pages={2658--2669},
  year={2022},
  publisher={ACS Publications}
}

@article{liu2026fgbench,
  title={Fgbench: A dataset and benchmark for molecular property reasoning at functional group-level in large language models},
  author={Liu, Xuan and Ouyang, Siru and Zhong, Xianrui and Han, Jiawei and Zhao, Huimin},
  journal={Advances in Neural Information Processing Systems},
  volume={38},
  year={2026}
}

@inproceedings{jang2025improving,
  title={Improving chemical understanding of llms via smiles parsing},
  author={Jang, Yunhui and Kim, Jaehyung and Ahn, Sungsoo},
  booktitle={Proceedings of the 2025 Conference on Empirical Methods in Natural Language Processing},
  pages={15694--15709},
  year={2025}
}

@article{cai2025mollangbench,
  title={Mollangbench: A comprehensive benchmark for language-prompted molecular structure recognition, editing, and generation},
  author={Cai, Feiyang and Bai, Jiahui and Tang, Tao and He, Guijuan and Luo, Joshua and Zhu, Tianyu and Pilla, Srikanth and Li, Gang and Liu, Ling and Luo, Feng},
  journal={arXiv preprint arXiv:2505.15054},
  year={2025}
}

@book{smith1911hindu,
  title={The hindu-arabic numerals},
  author={Smith, David Eugene and Karpinski, Louis Charles},
  year={1911},
  publisher={Ginn}
}

@article{kim2023pubchem,
  title={PubChem 2023 update},
  author={Kim, Sunghwan and Chen, Jie and Cheng, Tiejun and Gindulyte, Asta and He, Jia and He, Siqian and Li, Qingliang and Shoemaker, Benjamin A and Thiessen, Paul A and Yu, Bo and others},
  journal={Nucleic acids research},
  volume={51},
  number={D1},
  pages={D1373--D1380},
  year={2023},
  publisher={Oxford University Press}
}

@article{reiser2022graph,
  title={Graph neural networks for materials science and chemistry},
  author={Reiser, Patrick and Neubert, Marlen and Eberhard, Andr{\'e} and Torresi, Luca and Zhou, Chen and Shao, Chen and Metni, Houssam and van Hoesel, Clint and Schopmans, Henrik and Sommer, Timo and others},
  journal={Communications Materials},
  volume={3},
  number={1},
  pages={93},
  year={2022},
  publisher={Nature Publishing Group UK London}
}

@article{guo2025deepseek,
  title={{DeepSeek-R1} incentivizes reasoning in LLMs through reinforcement learning},
  author={Guo, Daya and Yang, Dejian and Zhang, Haowei and Song, Junxiao and Wang, Peiyi and Zhu, Qihao and Xu, Runxin and Zhang, Ruoyu and Ma, Shirong and Bi, Xiao and others},
  journal={Nature},
  volume={645},
  number={8081},
  pages={633--638},
  year={2025},
  publisher={Nature Publishing Group UK London}
}

@article{gemini2.5,
  title={Gemini 2.5: Pushing the frontier with advanced reasoning, multimodality, long context, and next generation agentic capabilities},
  author={Comanici, Gheorghe and Bieber, Eric and Schaekermann, Mike and Pasupat, Ice and Sachdeva, Noveen and Dhillon, Inderjit and Blistein, Marcel and Ram, Ori and Zhang, Dan and Rosen, Evan and others},
  journal={arXiv preprint arXiv:2507.06261},
  year={2025}
}

@misc{gemini3,
    title={Gemini 3 Pro Model Card},
    author={Google Deepmind},
    url={https://storage.googleapis.com/deepmind-media/Model-Cards/Gemini-3-Pro-Model-Card.pdf},
    year={2025}
}

@inproceedings{kusner2017grammar,
  title={Grammar variational autoencoder},
  author={Kusner, Matt J and Paige, Brooks and Hern{\'a}ndez-Lobato, Jos{\'e} Miguel},
  booktitle={International conference on machine learning},
  pages={1945--1954},
  year={2017},
  organization={PMLR}
}

@inproceedings{jin2018junction,
  title={Junction tree variational autoencoder for molecular graph generation},
  author={Jin, Wengong and Barzilay, Regina and Jaakkola, Tommi},
  booktitle={International conference on machine learning},
  pages={2323--2332},
  year={2018},
  organization={PMLR}
}

@article{molt5,
  title={Translation between molecules and natural language},
  author={Edwards, Carl and Lai, Tuan and Ros, Kevin and Honke, Garrett and Cho, Kyunghyun and Ji, Heng},
  journal={arXiv preprint arXiv:2204.11817},
  year={2022}
}

@inproceedings{pei2023biot5,
  title={{BioT5}: Enriching Cross-modal Integration in Biology with Chemical Knowledge and Natural Language Associations},
  author={Pei, Qizhi and Zhang, Wei and Zhu, Jinhua and Wu, Kehan and Gao, Kaiyuan and Wu, Lijun and Xia, Yingce and Yan, Rui},
  booktitle={The 2023 Conference on Empirical Methods in Natural Language Processing},
  year={2023},
}

@inproceedings{graphmvp,
  title={Pre-training Molecular Graph Representation with 3D Geometry},
  author={Liu, Shengchao and Wang, Hanchen and Liu, Weiyang and Lasenby, Joan and Guo, Hongyu and Tang, Jian},
  booktitle={International Conference on Learning Representations},
  year={2022},
}

@article{SMILES,
  title={{SMILES}, a chemical language and information system. 1. Introduction to methodology and encoding rules},
  author={Weininger, David},
  journal={Journal of Chemical Information and Computer Sciences},
  volume={28},
  number={1},
  pages={31--36},
  year={1988},
  publisher={ACS Publications}
}

@inproceedings{gilmer2017neural,
  title={Neural message passing for quantum chemistry},
  author={Gilmer, Justin and Schoenholz, Samuel S and Riley, Patrick F and Vinyals, Oriol and Dahl, George E},
  booktitle={International Conference on Machine Learning},
  pages={1263--1272},
  year={2017},
  organization={PMLR}
}

@inproceedings{kipf2017semi,
  title={Semi-supervised classification with graph convolutional networks},
  author={Kipf, Thomas N and Welling, Max},
  booktitle={International Conference on Learning Representations},
  year={2017}
}

@article{guo2023gpt_chemistry,
  title={What indeed can {GPT} models do in chemistry? {A} comprehensive benchmark on eight tasks},
  author={Guo, Taicheng and Guo, Kehan and Nan, Bowen and Liang, Zhenwen and Guo, Zhichun and Chawla, Nitesh and Wiest, Olaf and Zhang, Xiangliang},
  journal={Advances in Neural Information Processing Systems},
  volume={36},
  year={2023}
}

@article{GPT5,
  title={Openai gpt-5 system card},
  author={Singh, Aaditya and Fry, Adam and Perelman, Adam and Tart, Adam and Ganesh, Adi and El-Kishky, Ahmed and McLaughlin, Aidan and Low, Aiden and Ostrow, AJ and Ananthram, Akhila and others},
  journal={arXiv preprint arXiv:2601.03267},
  year={2025}
}

@article{ChemBench,
  title={A framework for evaluating the chemical knowledge and reasoning abilities of large language models against the expertise of chemists},
  author={Mirza, Adrian and Alampara, Nawaf and Kunchapu, Sreekanth and R{\'\i}os-Garc{\'\i}a, Marti{\~n}o and Emoekabu, Benedict and Krishnan, Aswanth and Gupta, Tanya and Schilling-Wilhelmi, Mara and Okereke, Macjonathan and Aneesh, Anagha and others},
  journal={Nature Chemistry},
  volume={17},
  number={7},
  pages={1027--1034},
  year={2025},
  publisher={Nature Publishing Group UK London}
}

@inproceedings{liu2023molca,
  title={Molca: Molecular graph-language modeling with cross-modal projector and uni-modal adapter},
  author={Liu, Zhiyuan and Li, Sihang and Luo, Yanchen and Fei, Hao and Cao, Yixin and Kawaguchi, Kenji and Wang, Xiang and Chua, Tat-Seng},
  booktitle={Proceedings of the 2023 Conference on Empirical Methods in Natural Language Processing},
  pages={15623--15638},
  year={2023}
}

@inproceedings{schutt2018schnet,
  title={{SchNet}: A continuous-filter convolutional neural network for modeling quantum interactions},
  author={Sch{\"u}tt, Kristof T and Kindermans, Pieter-Jan and Sauceda, Huziel E and Chmiela, Stefan and Tkatchenko, Alexandre and M{\"u}ller, Klaus-Robert},
  booktitle={Advances in Neural Information Processing Systems},
  volume={30},
  year={2018}
}

@misc{rdkit,
  title={{RDKit}: Open-source cheminformatics},
  author={Landrum, Greg and others},
  howpublished={\url{http://www.rdkit.org}},
  note={Version 2023.09},
  year={2006}
}

@article{SELFIES,
  title={Self-referencing embedded strings (SELFIES): A 100\% robust molecular string representation},
  author={Krenn, Mario and H{\"a}se, Florian and Nigam, AkshatKumar and Friederich, Pascal and Aspuru-Guzik, Alan},
  journal={Machine Learning: Science and Technology},
  volume={1},
  number={4},
  pages={045024},
  year={2020},
  publisher={IOP Publishing}
}

@article{t_SMILES,
  title={t-SMILES: a fragment-based molecular representation framework for de novo ligand design},
  author={Wu, Juan-Ni and Wang, Tong and Chen, Yue and Tang, Li-Juan and Wu, Hai-Long and Yu, Ru-Qin},
  journal={Nature Communications},
  volume={15},
  number={1},
  pages={4993},
  year={2024},
  publisher={Nature Publishing Group UK London}
}

@inproceedings{fang2025molparser,
  title={MolParser: End-to-end visual recognition of molecule structures in the wild},
  author={Fang, Xi and Wang, Jiankun and Cai, Xiaochen and Chen, Shangqian and Yang, Shuwen and Tao, Haoyi and Wang, Nan and Yao, Lin and Zhang, Linfeng and Ke, Guolin},
  booktitle={Proceedings of the IEEE/CVF International Conference on Computer Vision},
  pages={24528--24538},
  year={2025}
}

@article{PSMILES,
  title={polyBERT: a chemical language model to enable fully machine-driven ultrafast polymer informatics},
  author={Kuenneth, Christopher and Ramprasad, Rampi},
  journal={Nature communications},
  volume={14},
  number={1},
  pages={4099},
  year={2023},
  publisher={Nature Publishing Group UK London}
}

@article{li2026programmable,
  title={Programmable divergent electrochemical ring-opening multifunctionalization of strained rings},
  author={Li, Yajuan and Lang, Yatao and He, Shu-Fan and Li, Daixi and Liu, Ke-Xin and Ai, Wenying and Jiang, Yong and Zhu, Chen and Shen, Tao},
  journal={Nature Chemistry},
  pages={1--13},
  year={2026},
  publisher={Nature Publishing Group UK London}
}

@inproceedings{molinstructions,
  title={Mol-instructions: A large-scale biomolecular instruction dataset for large language models},
  author={Fang, Yin and Liang, Xiaozhuan and Zhang, Ningyu and Liu, Kangwei and Huang, Rui and Chen, Zhuo and Fan, Xiaohui and Chen, Huajun},
  booktitle={International Conference on Learning Representations},
  volume={2024},
  pages={48221--48251},
  year={2024}
}

@inproceedings{cao2025instructmol,
  title={Instructmol: Multi-modal integration for building a versatile and reliable molecular assistant in drug discovery},
  author={Cao, He and Liu, Zijing and Lu, Xingyu and Yao, Yuan and Li, Yu},
  booktitle={Proceedings of the 31st International Conference on Computational Linguistics},
  pages={354--379},
  year={2025}
}

@article{lv2024navigating,
  title={Navigating chemical-linguistic sharing space with heterogeneous molecular encoding},
  author={Lv, Liuzhenghao and Li, Hao and Wang, Yu and Yan, Zhiyuan and Chen, Zijun and Lin, Zongying and Yuan, Li and Tian, Yonghong},
  journal={arXiv preprint arXiv:2412.20888},
  year={2024}
}

@article{skinnider2024invalid,
  title={Invalid SMILES are beneficial rather than detrimental to chemical language models},
  author={Skinnider, Michael A},
  journal={Nature Machine Intelligence},
  volume={6},
  number={4},
  pages={437--448},
  year={2024},
  publisher={Nature Publishing Group UK London}
}

@inproceedings{liu2023molxpt,
  title={Molxpt: Wrapping molecules with text for generative pre-training},
  author={Liu, Zequn and Zhang, Wei and Xia, Yingce and Wu, Lijun and Xie, Shufang and Qin, Tao and Zhang, Ming and Liu, Tie-Yan},
  booktitle={Proceedings of the 61st Annual Meeting of the Association for Computational Linguistics (Volume 2: Short Papers)},
  pages={1606--1616},
  year={2023}
}

@inproceedings{hao2025detect,
  title={How to Detect and Defeat Molecular Mirage: A Metric-Driven Benchmark for Hallucination in LLM-based Molecular Comprehension},
  author={Hao, Li and Lv, Liuzhenghao and Liu, Zijing and Yan, Zhiyuan and Wang, Yu and Tian, Yonghong and Li, Yu and Yuan, Li and others},
  booktitle={NeurIPS 2025 AI for Science Workshop},
  year={2025}
}

@inproceedings{lyy-blending2d3d,
title={Multimodal Molecular Pretraining via Modality Blending},
author={Qiying Yu and Yudi Zhang and Yuyan Ni and Shikun Feng and Yanyan Lan and Hao Zhou and Jingjing Liu},
booktitle={The Twelfth International Conference on Learning Representations},
year={2024},
url={https://openreview.net/forum?id=oM7Jbxdk6Z}
}

@article{lyy-unicorn2d3d,
  title={UniCorn: A Unified Contrastive Learning Approach for Multi-view Molecular Representation Learning},
  author={Feng, Shikun and Ni, Yuyan and Li, Minghao and Huang, Yanwen and Ma, Zhi-Ming and Ma, Wei-Ying and Lan, Yanyan},
  journal={arXiv preprint arXiv:2405.10343},
  year={2024}
}

@inproceedings{seidl2023clamp1,
  title={Enhancing activity prediction models in drug discovery with the ability to understand human language},
  author={Seidl, Philipp and Vall, Andreu and Hochreiter, Sepp and Klambauer, G{\"u}nter},
  booktitle={International Conference on Machine Learning},
  pages={30458--30490},
  year={2023},
  organization={PMLR}
}

@article{momu,
  title={A molecular multimodal foundation model associating molecule graphs with natural language},
  author={Su, Bing and Du, Dazhao and Yang, Zhao and Zhou, Yujie and Li, Jiangmeng and Rao, Anyi and Sun, Hao and Lu, Zhiwu and Wen, Ji-Rong},
  journal={arXiv preprint arXiv:2209.05481},
  year={2022}
}

@article{kim2026mol-llama,
  title={Mol-llama: Towards general understanding of molecules in large molecular language model},
  author={Kim, Dongki and Lee, Wonbin and Hwang, Sung Ju},
  journal={Advances in Neural Information Processing Systems},
  volume={38},
  pages={26921--26960},
  year={2026}
}

@inproceedings{wu2025molerr2fix,
  title={MolErr2Fix: Benchmarking LLM Trustworthiness in Chemistry via Modular Error Detection, Localization, Explanation, and Correction},
  author={Wu, Yuyang and Ye, Jinhui and Zhang, Shuhao and Dai, Lu and Bisk, Yonatan and Isayev, Olexandr},
  booktitle={Proceedings of the 2025 Conference on Empirical Methods in Natural Language Processing},
  pages={19365--19382},
  year={2025}
}

@inproceedings{zhouunimol,
  title={Uni-Mol: A Universal 3D Molecular Representation Learning Framework},
  author={Zhou, Gengmo and Gao, Zhifeng and Ding, Qiankun and Zheng, Hang and Xu, Hongteng and Wei, Zhewei and Zhang, Linfeng and Ke, Guolin},
  booktitle={The Eleventh International Conference on Learning Representations},
  year={2023}
}

@article{hafidi2020graphcl,
  title={Graphcl: Contrastive self-supervised learning of graph representations},
  author={Hafidi, Hakim and Ghogho, Mounir and Ciblat, Philippe and Swami, Ananthram},
  journal={arXiv preprint arXiv:2007.08025},
  year={2020}
}

@article{kv-plm,
  title={Multi-modal molecule structure--text model for text-based retrieval and editing},
  author={Liu, Shengchao and Nie, Weili and Wang, Chengpeng and Lu, Jiarui and Qiao, Zhuoran and Liu, Ling and Tang, Jian and Xiao, Chaowei and Anandkumar, Animashree},
  journal={Nature Machine Intelligence},
  volume={5},
  number={12},
  pages={1447--1457},
  year={2023},
  publisher={Nature Publishing Group UK London}
}

@article{wei2022cot,
  title={Chain-of-thought prompting elicits reasoning in large language models},
  author={Wei, Jason and Wang, Xuezhi and Schuurmans, Dale and Bosma, Maarten and Xia, Fei and Chi, Ed and Le, Quoc V and Zhou, Denny and others},
  journal={Advances in neural information processing systems},
  volume={35},
  pages={24824--24837},
  year={2022}
}

@inproceedings{li20243dmolm,
  title={Towards 3d molecule-text interpretation in language models},
  author={Li, Sihang and Liu, Zhiyuan and Luo, Yanchen and Wang, Xiang and He, Xiangnan and Kawaguchi, Kenji and Chua, Tat-Seng and Tian, Qi},
  booktitle={International Conference on Learning Representations},
  volume={2024},
  pages={17352--17371},
  year={2024}
}

@article{zhong2022root,
  title={Root-aligned SMILES: a tight representation for chemical reaction prediction},
  author={Zhong, Zipeng and Song, Jie and Feng, Zunlei and Liu, Tiantao and Jia, Lingxiang and Yao, Shaolun and Wu, Min and Hou, Tingjun and Song, Mingli},
  journal={Chemical Science},
  volume={13},
  number={31},
  pages={9023--9034},
  year={2022},
  publisher={Royal Society of Chemistry}
}

@misc{wo2026055477,
  title = {Macrocyclic Compounds as Modulators of KRAS and Uses Thereof},
  author = {WURZ, Ryan Paul and YAMANO, Michael Masaharu and SARDINI JR., Stephen and ZHAO, Wei and TANWAR, Lalita and LI, Yunxiao and LANMAN, Brian Alan and AMEGADZIE, Albert K.},
  number = {WO2026055477},
  type = {WO Patent},
  url = {https://patentscope.wipo.int/search/en/WO2026055477}
}

@inproceedings{yu2024llasmol,
  title={LlaSMol: Advancing Large Language Models for Chemistry with a Large-Scale, Comprehensive, High-Quality Instruction Tuning Dataset},
  author={Yu, Botao and Baker, Frazier N and Chen, Ziqi and Ning, Xia and Sun, Huan},
  booktitle={ACL 2024 Workshop Language+ Molecules},
  year={2024}
}

@article{ChemIQ,
  title={Assessing the chemical intelligence of large language models},
  author={Runcie, Nicholas T and Deane, Charlotte M and Imrie, Fergus},
  journal={Journal of Chemical Information and Modeling},
  volume={66},
  number={1},
  pages={216--227},
  year={2025},
  publisher={ACS Publications}
}

@article{ChemCoTBench,
  title={Beyond chemical qa: Evaluating llm's chemical reasoning with modular chemical operations},
  author={Hao, Li and Cao, He and Feng, Bin and Shao, Daniel and Tang, Robert and Yan, Zhiyuan and Tian, Yonghong and Yuan, Li and Li, Yu},
  journal={Advances in Neural Information Processing Systems},
  volume={38},
  year={2026}
}

@article{chen2024large,
  title={A large-scale reaction dataset of mechanistic pathways of organic reactions},
  author={Chen, Shuan and Babazade, Ramil and Kim, Taewan and Han, Sunkyu and Jung, Yousung},
  journal={Scientific Data},
  volume={11},
  number={1},
  pages={863},
  year={2024},
  publisher={Nature Publishing Group UK London}
}
